\documentclass[letterpaper,journal]{IEEEtran}
\usepackage{amsmath,amsfonts}
\usepackage{algorithmic}
\usepackage{algorithm}
\usepackage{array}
\usepackage[caption=false,font=normalsize,labelfont=rm,textfont=rm]{subfig}
\usepackage{textcomp}
\usepackage{stfloats}
\usepackage{url}
\usepackage{verbatim}
\usepackage{cite}
\usepackage{multirow}
\usepackage{graphicx}
\usepackage{xcolor}
\usepackage{gensymb}
\usepackage{amssymb}
\begin{document}

\title{A Compact Omnidirectional Meanderline Antenna Array for Wireless Security Using Dynamic Magnitude and Phase Pattern Modulation}

\author{Sheng Huang,~\IEEEmembership{Member,~IEEE}, Jacob R. Randall,~\IEEEmembership{Student Member,~IEEE}, Cory Hilton,~\IEEEmembership{Student Member,~IEEE}, Jeffrey A. Nanzer, \IEEEmembership{Senior Member,~IEEE}

\thanks{Manuscript received June 13, 2025. \textit{(Corresponding author: Jeffrey A. Nanzer.)}
} 
\thanks{The authors are with the Department of Electrical and Computer Engineering, Michigan State University, East Lansing, MI 48824 USA (e-mail:
	huang287@msu.edu; randa130@msu.edu; hiltonc2@msu.edu; nanzer@msu.edu).}
}
\markboth{Journal of \LaTeX\ Class Files,~Vol.~14, No.~8, August~2021}%
{Shell \MakeLowercase{\textit{et al.}}: A Sample Article Using IEEEtran.cls for IEEE Journals}

\IEEEpubid{}

\maketitle

\begin{abstract}
A compact dynamic four-element array with omnidirectional H-plane coverage is presented for planar physical-layer security using antenna-level directional modulation. The proposed approach achieves angularly selective information transmission without phased-array beamforming or multiple RF chains by dynamically switching the excitation paths of a four-element array. The antenna comprises four printed meander-line monopole elements operating at 5.05~GHz with independently controlled differential power excitation, which introduces magnitude and phase pattern modulation, leading to dynamic motion of the apparent element spacing, resulting in strongly angle-dependent signal distortion and bit error rate (BER) performance. Reliable information recovery is confined to a narrow broadside region in the E-plane, while significantly elevated BER is observed at off-broadside angles. In contrast, the H-plane radiation remains static and omnidirectional, enabling full $360^{\circ}$ information-recoverable coverage in the orthogonal plane. The antenna is fabricated on a single-layer Rogers RO4350B substrate with a compact footprint of $0.55 \times 1.73\lambda_{0}^{2}$. A four-path switching network implemented using commercial RF components validates the concept experimentally. Communication measurements under high-SNR conditions ($>19$~dB) using 16-QAM demonstrate a planar information beamwidth (IB) below $24^{\circ}$, confirming effective antenna-level directional modulation with angle-dependent BER characteristics and omnidirectional H-plane coverage.

\end{abstract}


\section{Introduction}
\IEEEPARstart{S}{patially} selective secure transmission is increasingly desirable in compact wireless platforms that must preserve broad angular coverage while limiting information recovery to a narrow region of space. This requirement appears in low-profile access points, distributed sensors, embedded terminals, and lightweight autonomous nodes, where the radio front end must remain simple and power efficient. Under these conditions, physical-layer security (PLS) offers an attractive complement to higher-layer encryption by embedding confidentiality into the electromagnetic transmission process itself \cite{5751298,7467419,8509094}. Directional modulation (DM) is one of the most representative antenna-based PLS techniques because the transmitted constellation is intentionally made angle dependent: the intended receiver observes an undistorted symbol set, whereas off-axis observers experience constellation deformation and increased bit error rate (BER) \cite{5159486,5422702,6645431}.
\begin{figure}[t]
\begin{center}
\includegraphics[width=2.7in]{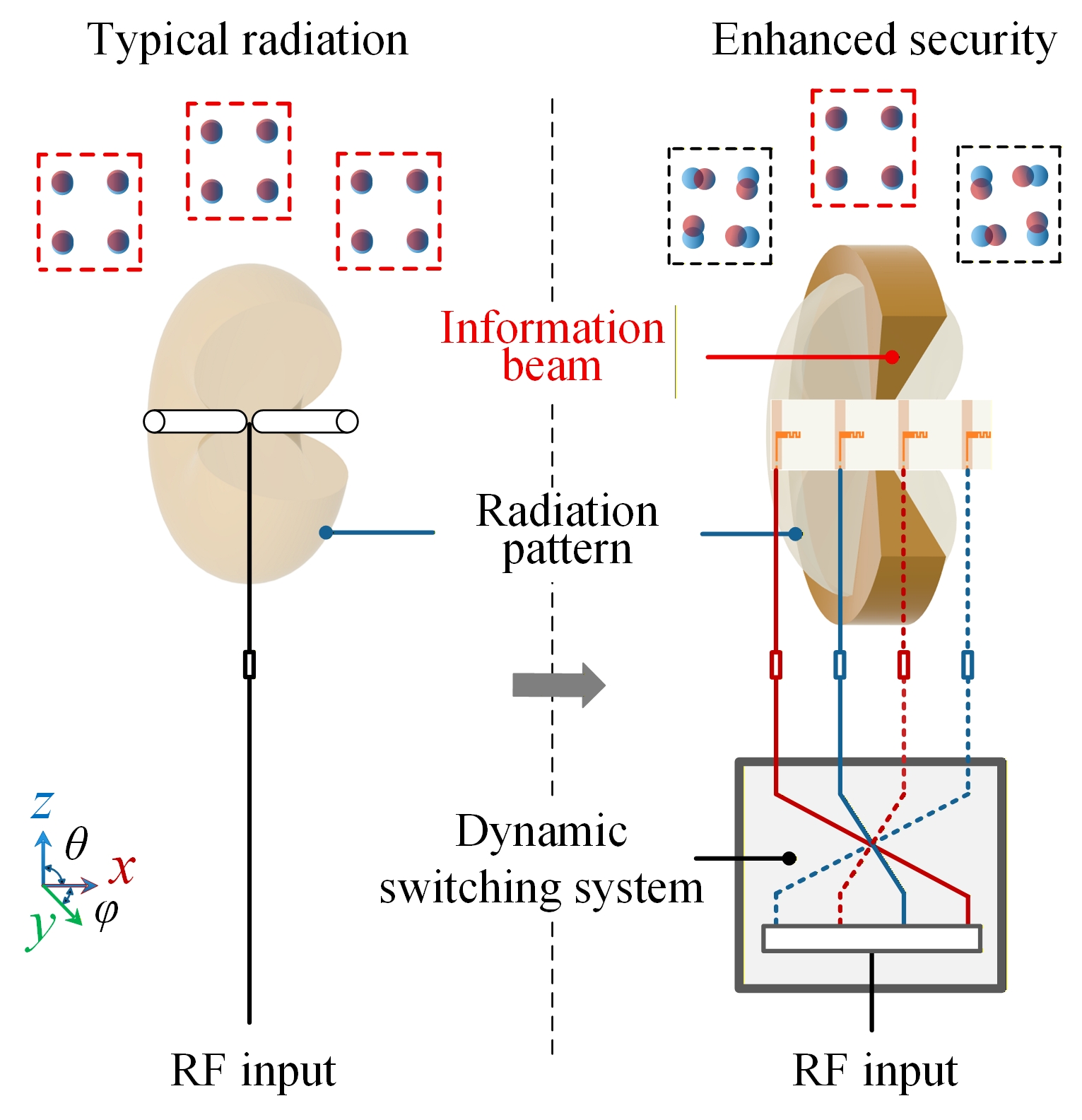}%
\end{center}
\caption{Topology of improved information security using the proposed dynamic array. Typical radiation from small antennas results in a wide information-recoverable region. The proposed dynamic array enables spatially selective information transmission by introducing differential magnitude and phase variations in the complex radiation patterns, such that only receivers located within a narrow sector in the $\varphi = 0^\circ$ plane can successfully recover the modulated signal. In contrast, receivers in other directions are unable to retrieve the information, thereby defining unrecoverable regions. Meanwhile, in the $\varphi = 90^\circ$ plane, the antenna maintains omnidirectional radiation without directional modulation, allowing information to be recovered across all angles. This spatial modulation scheme enhances planar physical-layer security.}\label{overview}
\end{figure}
Classical DM implementations are predominantly based on phased arrays and related multi-aperture transmitters \cite{5159486,5422702,5439878,6544472,9184842}. These architectures provide flexible spatial control, but they typically require multiple coherent RF chains, calibration-sensitive feed networks, and a physical footprint that is difficult to reconcile with compact planar hardware. To reduce this overhead, alternative antenna-level solutions have been explored, including near-field direct antenna modulation \cite{4684619}, distributed dynamic arrays \cite{9665259}, single-element dynamic antennas \cite{10161710}, electrically small dynamic DM antennas \cite{9674846}, and phase-center control through spatial amplitude modulation \cite{10286341}. Despite this progress, simultaneously achieving compactness, planar manufacturability, omnidirectional coverage, and a narrow information-recoverable sector remains challenging.

The present work addresses this challenge through a compact four-element dynamic array in which omnidirectional radiation and security selectivity are intentionally separated between the two principal planes. As illustrated in Fig.~\ref{overview}, the proposed antenna maintains omnidirectional behavior in the $\varphi=90^\circ$ plane, while switching-induced magnitude and phase perturbations create a confined information beam in the $\varphi=0^\circ$ plane. This operating principle is fundamentally different from conventional beam-steering arrays: instead of continuously synthesizing a desired beam with multiple phase-controlled channels, the proposed system uses a single RF input and a low-complexity switching network to alternate among dynamic array states that reshape the complex radiation pattern.
The array is implemented using printed meander-line monopole elements. Meander-line radiators are widely adopted for miniaturization because the folded current path increases the effective electrical length without enlarging the footprint \cite{99054,6171817}. Their compatibility with printed fabrication has supported extensive use in compact wireless devices, wearable systems, RFID structures, and system-in-package antennas \cite{1504825,9263312,9128053,6335459,5299014,6648416,9591355,7748548,7384427,5979187,6843861,8758307}. Here, that miniaturization capability is exploited not simply to reduce size, but to realize a fully planar four-element dynamic aperture whose total electrical size remains only $0.55 \times 1.73 \times 0.0083\,\lambda_{0}^{3}$ at 5 GHz.
\begin{figure}[t]
	\centering
\includegraphics[width=2in]{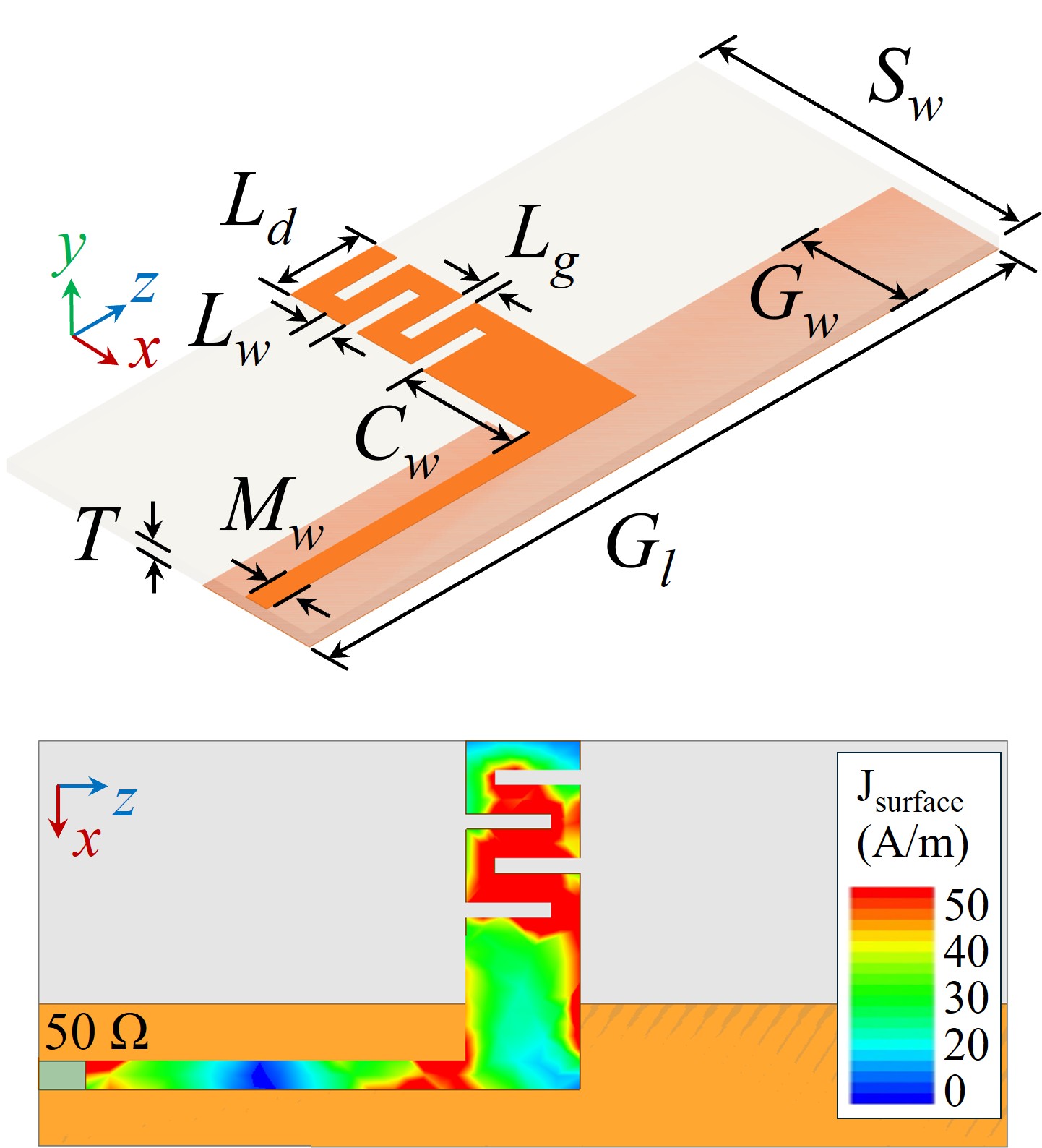}%
	\caption{Single-element planar meander-line antenna with microstrip feeding on Rogers 4350B and dimensions selected for resonance at 5~GHz. The structure consists of a meander-line radiating element, a capacitive patch, a microstrip feed, and an extended ground plane.}
\end{figure}
\begin{table}[t]
	\caption{Optimized Dimensions of the Planar Meander Line Antenna for the Resonance at 5 GHz   (Unit: mm)\label{tab:table1}}
	\centering
	\renewcommand{\arraystretch}{1.2}
	{
		\begin{tabular}{ccccccccc} 
			\hline \hline
			\(L_{g}\)&\(L_{w}\) &\(L_{d}\) & \(M_{w}\) &\(G_{w}\) &\(G_{l}\)&\(T\)&\(C_{w}\) &\(S_{w}\)
			\\  
			0.55 & 1& 4 & 1.08& 5 & 34 & 0.5 & 5& 14.2
			\\
			\hline\hline
		\end{tabular}
	}
\end{table}By switching the feed-path power ratios among the four radiators, the proposed array produces state-dependent E-plane magnitude and phase responses while leaving the H-plane quasi-static and omnidirectional. As a result, angular BER selectivity is obtained without sacrificing broad azimuthal coverage. A practical four-path RF switching system is implemented using commercial switches, phase shifters, attenuators, and power splitters, enabling a hardware-efficient realization suitable for compact front ends. Full-wave analysis and 16-QAM measurements confirm that the antenna forms a highly confined secure region near broadside while maintaining omnidirectional communication outside the modulated plane.
\begin{figure}[t]
	\centering
\subfloat[]{\includegraphics[width=3.2in]{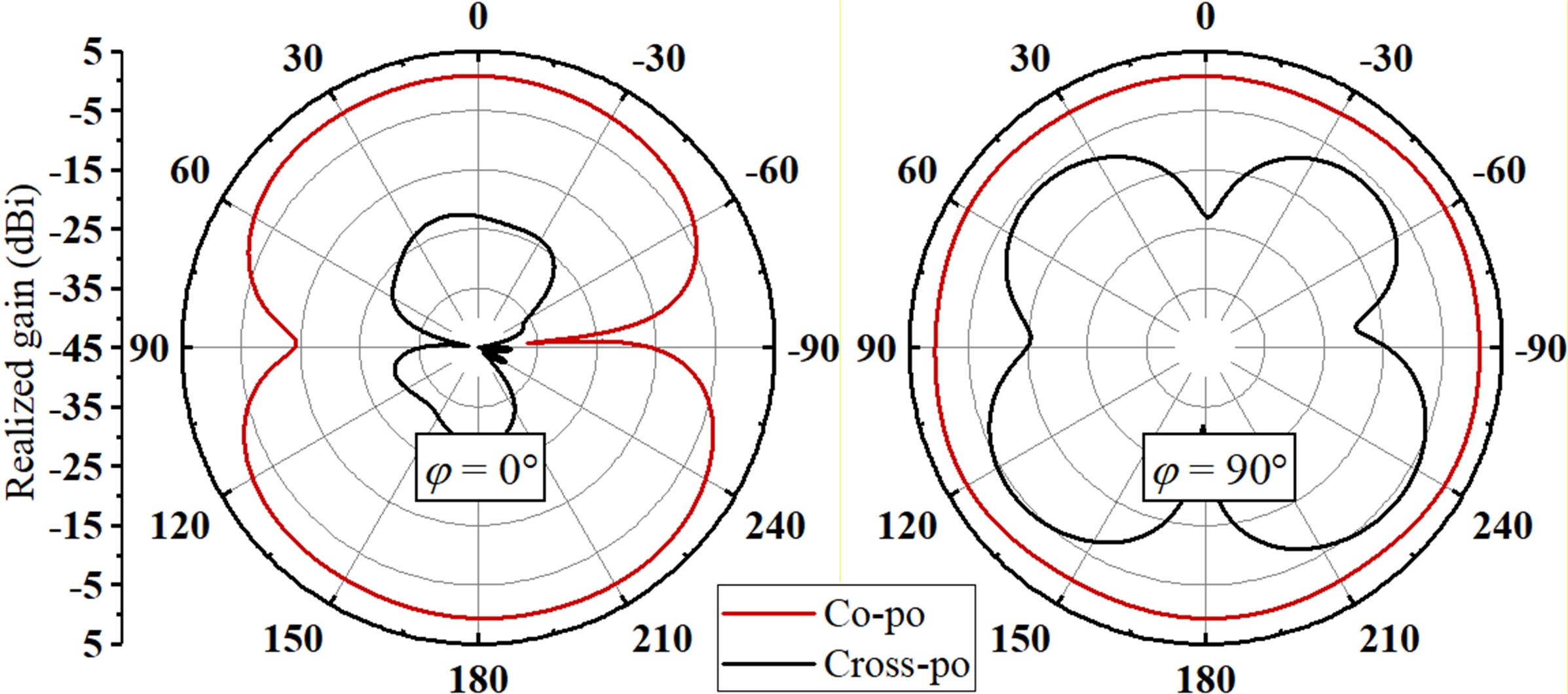}}\hfil
\subfloat[]{\includegraphics[width=1.7in]{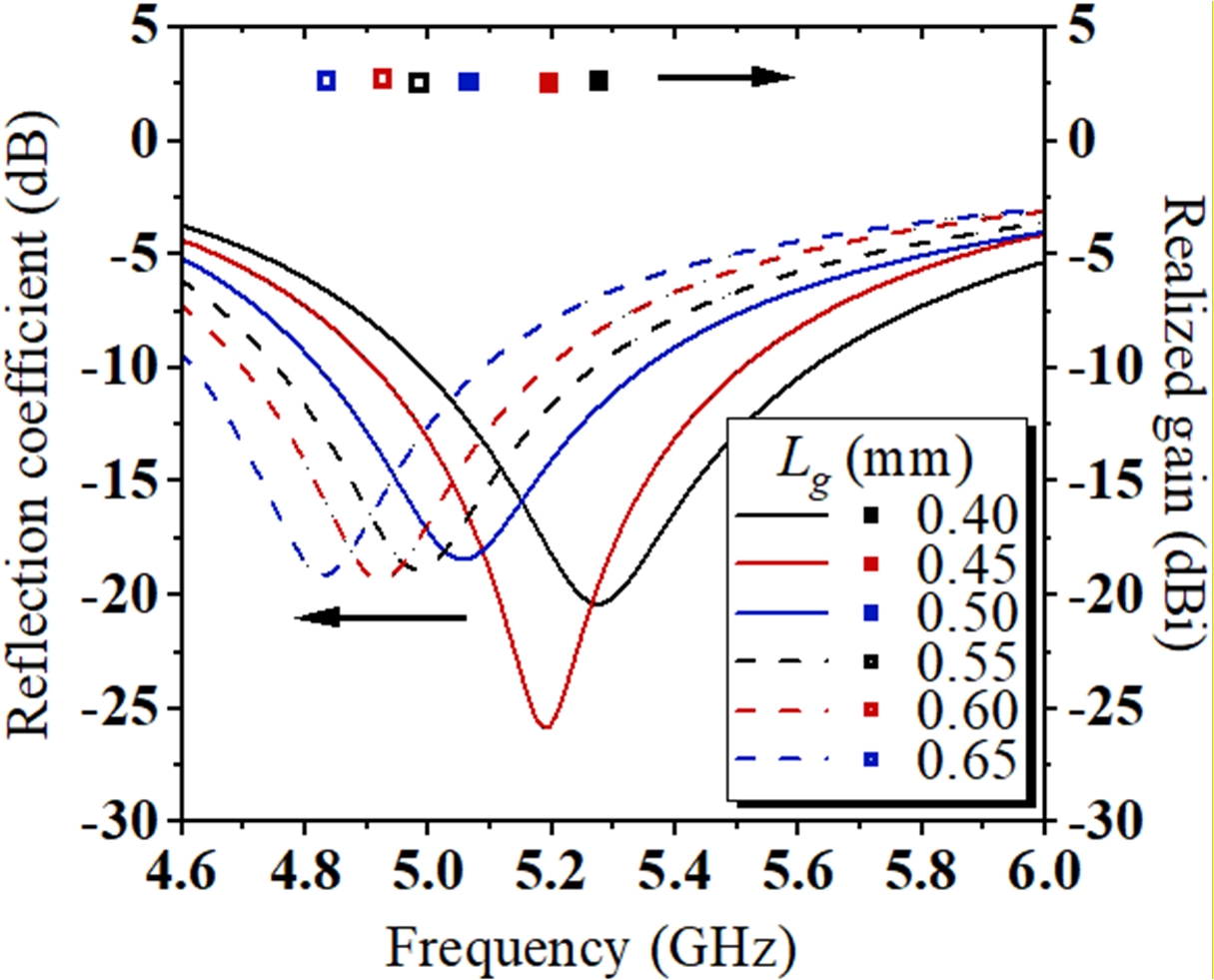}%
	}
	\hfil
\subfloat[]{\includegraphics[width=1.7in]{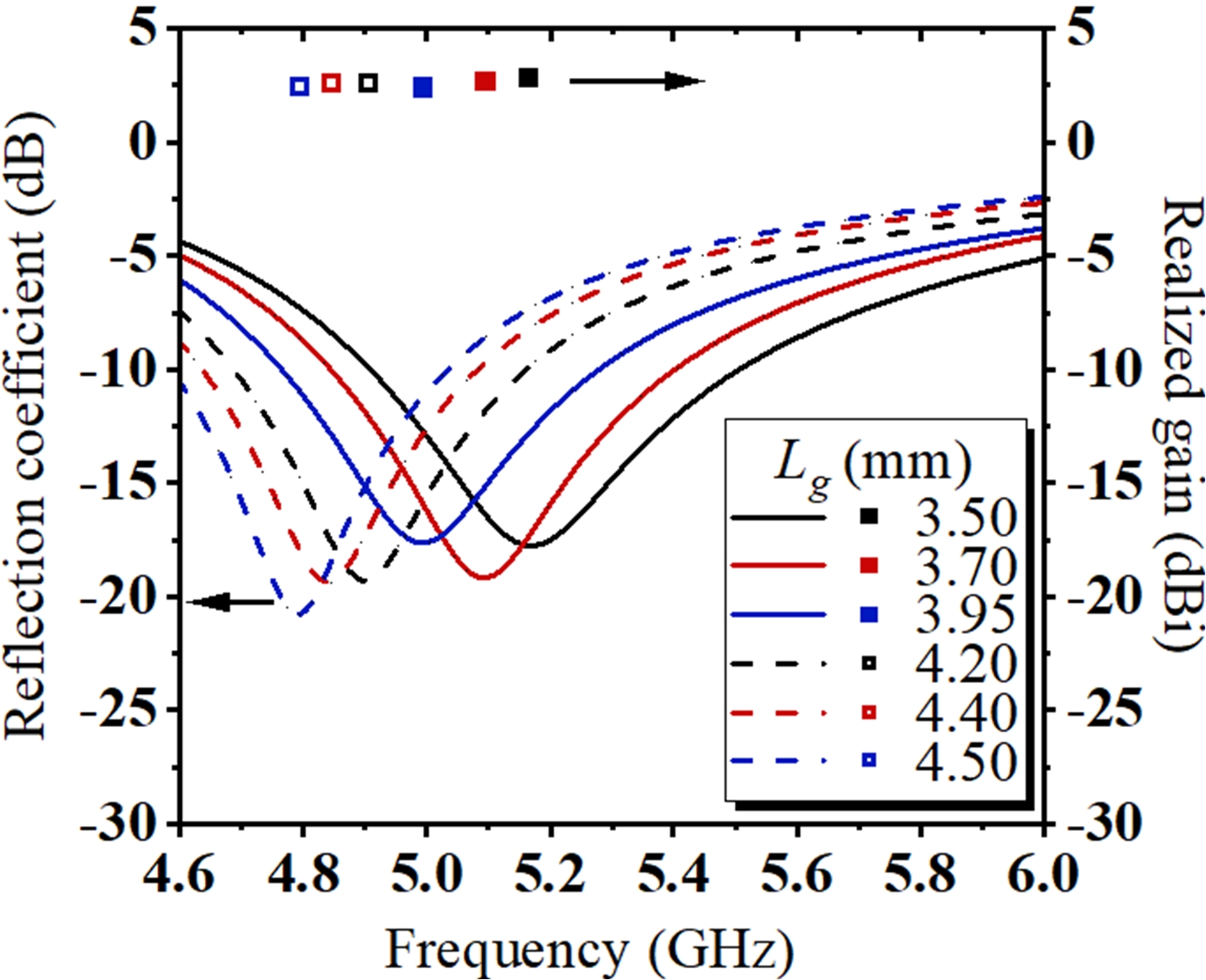}%
	}\hfil
	\caption{Simulated radiation performance and matching of the single element. (a) Co- and cross-polarized realized gain in the E- and H-planes at 5 GHz. Frequency tuning by (b) vertical length \(L_{g}\) and (c) antenna width \(L_{d}\).}\label{fig:single_element_tuning}
\end{figure}

The main contributions of this paper are summarized as follows. First, a compact printed four-element dynamic array is developed for planar physical-layer security with omnidirectional H-plane coverage. Second, the proposed architecture achieves directional information selectivity using a single-RF-input switching network rather than a conventional multi-channel phased-array transmitter. Third, the meander-line implementation enables a range of different dynamic aperture modulations through magnitude pattern dynamics, phase pattern dynamics, or both in the E-plane while maintaining omnidirectional information coverage in the H-plane. Finally, the concept is validated through simulation, hardware implementation, and communication measurements.
The remainder of this paper is organized as follows. Section II presents the geometry evolution from the single meander-line element to the proposed four-element dynamic array, together with the associated theoretical interpretation of the switching states and antenna-level modulation mechanism. Section III investigates the communication characteristics of the proposed array, including angular BER behavior and the effects of key design parameters. Section IV describes the fabricated prototype, the switching hardware, and the experimental validation of the proposed dynamic antenna. Section V concludes the paper.
\section{Planar Meander-Line Antenna Design}
\subsection{Single Meander-Line Monopole}

We initiate the design of the proposed dynamic array from a planar meander-line monopole, as depicted in Fig.~2. To enable low-cost and streamlined fabrication, the complete radiator and feed architecture, including the meandered trace, capacitive patch, microstrip feed, and ground plane, is implemented on a single-layer Rogers RO4350B substrate with relative permittivity $\varepsilon_r=3.48$ and loss tangent $\tan\delta=0.0037$. A substrate thickness $T=0.5~\mathrm{mm}$ is selected to maintain a compact and lightweight profile.
The antenna is excited by a $50~\Omega$ microstrip line of width $M_w$. A capacitive patch of length $C_w$, following the loading concept reported in~\cite{4020418}, is introduced to increase the effective electrical length and provide a practical planar feed transition, thereby facilitating miniaturization at the target frequency. The meander-line radiator comprises $N=2$ sections, where the strip width and spacing are denoted by $L_w$ and $L_g$, respectively. The capacitive patch and the meandered element share the same overall lateral dimension $L_d$.
\begin{figure}[t]
	\centering
\includegraphics[width=2.7in]{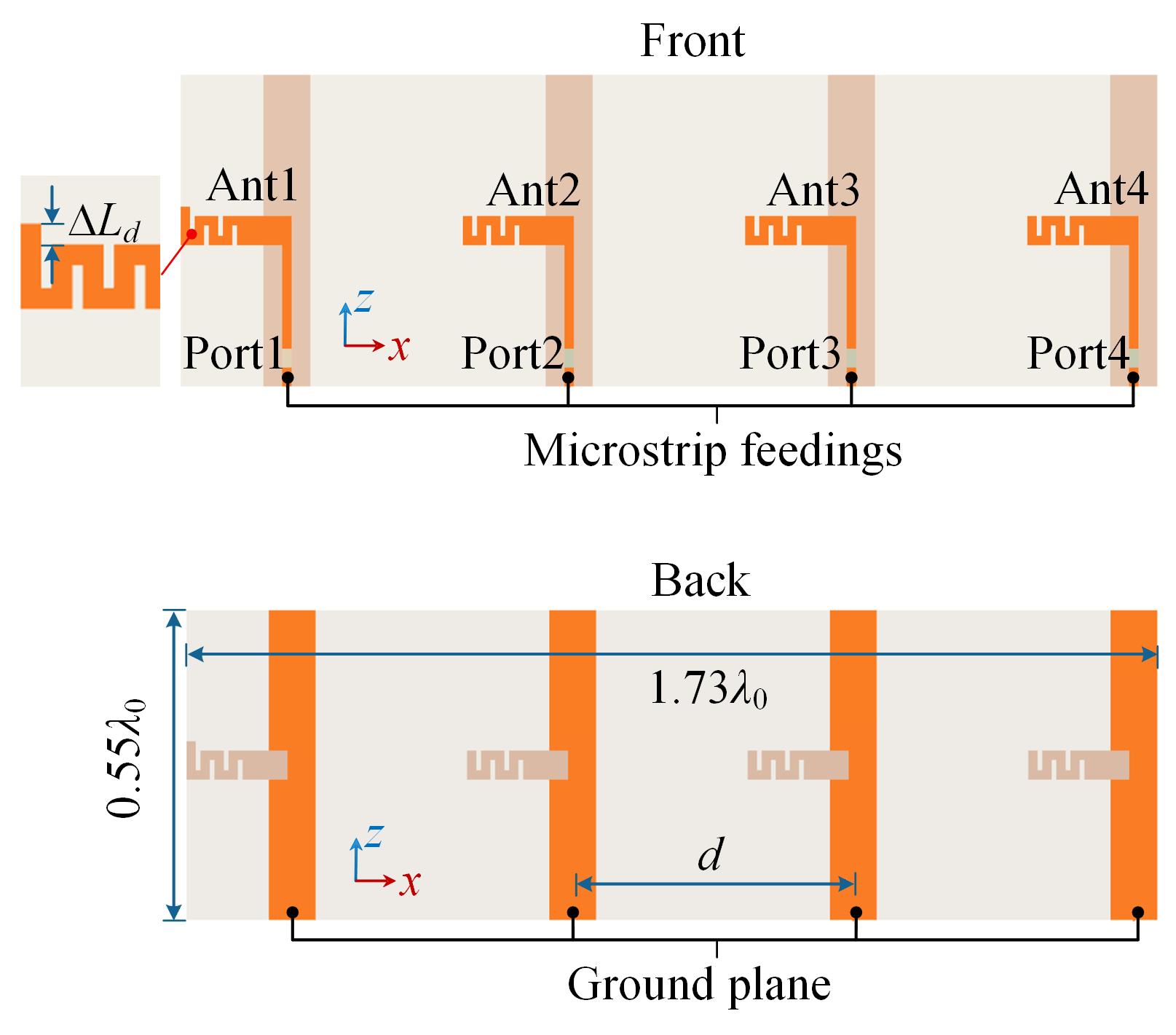}
	\caption{Structure of the designed dynamic four-element array using the proposed planar meander-line elements at 5~GHz.}\label{fig:four_element_structure}
\end{figure}
\begin{figure}[t]
	\centering
\subfloat[]{\includegraphics[width=3.2in]{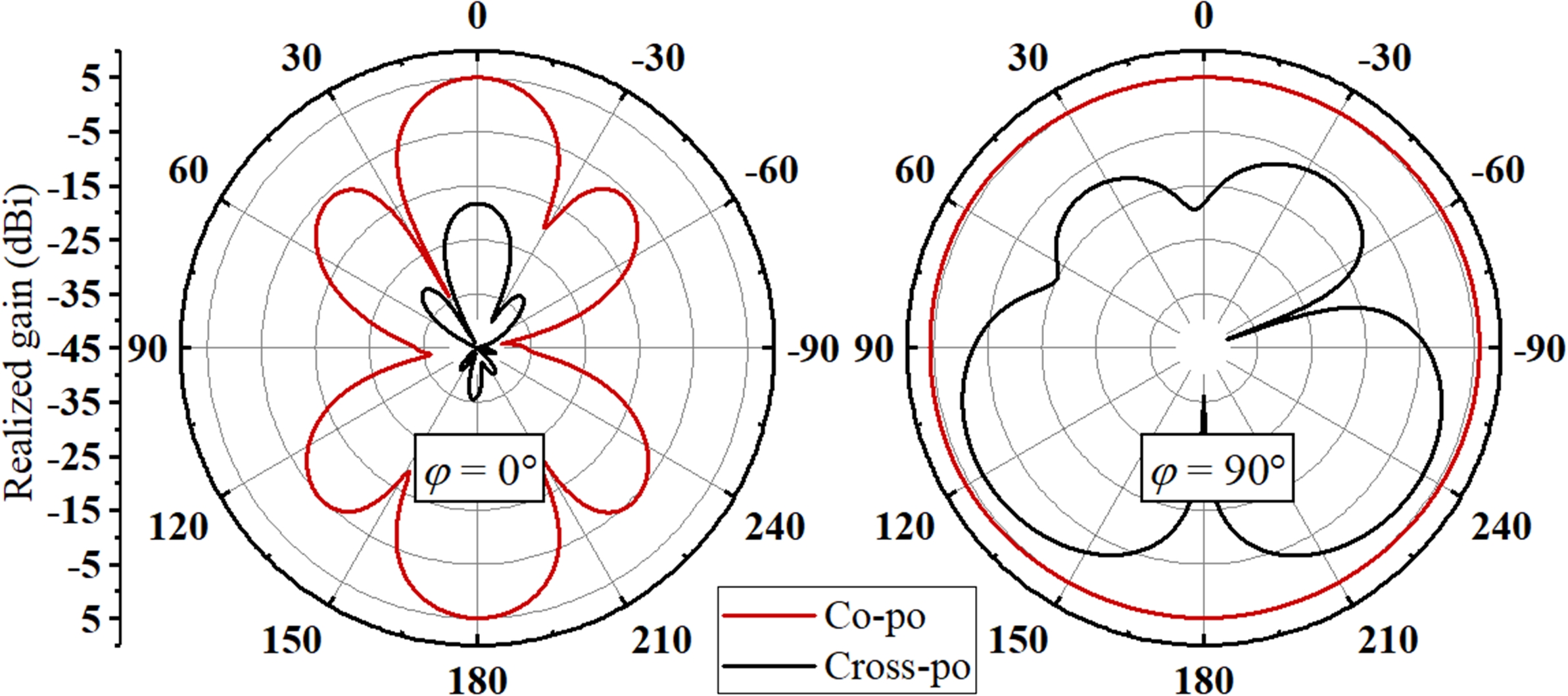}}\hfil
\subfloat[]{\includegraphics[width=3in]{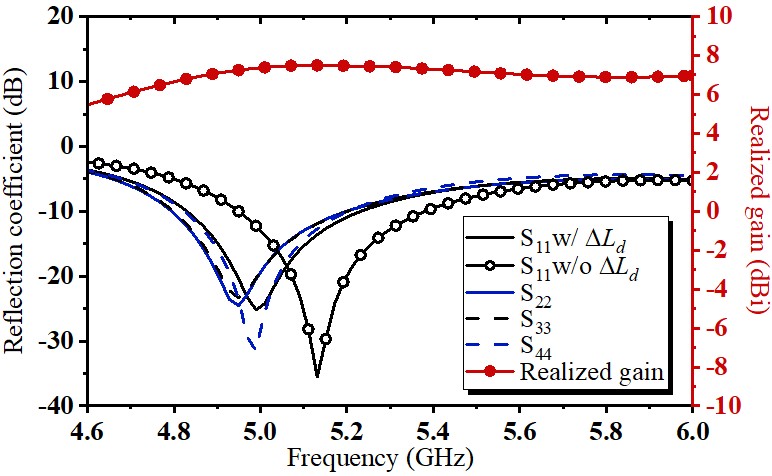}}
\hfil
\caption{Simulated radiation performance and matching of the four-element dynamic array. (a) Realized gain radiation patterns at 5 GHz. (b) Reflection coefficient and realized gain.}\label{fig:four_element_matching}
\end{figure}
To preserve omnidirectional radiation characteristics, the ground plane dimensions (length $G_l$ and width $G_w$) are jointly optimized so that the direction of maximum realized gain remains orthogonal to the electric-field polarization. All geometrical parameters are tuned for resonance at $5~\mathrm{GHz}$ and are summarized in Table~\ref{tab:table1}. The design process of the meander line is well developed in \cite{99054,10379428}. Self-resonance is obtained when the inductive reactance of the folded section cancels the capacitive reactance of the short radiator, i.e., \(X_C = X_D\), where \(X_C\) and \(X_D\) denote the inductive and capacitive reactances of the equivalent meander-line model, respectively. This condition provides the initial guideline for selecting the meander section number and folded path length. The final values of \(N\), \(L_w\), \(L_g\), and \(L_d\) are then refined in full-wave simulation to account for the printed planar geometry, finite ground plane, capacitive patch, and microstrip feed. 

The simulated surface current distribution obtained using ANSYS HFSS is shown in Fig.~2, which exhibits the characteristic behavior of a meander-line radiator: currents at successive folds are co-directed, whereas currents on adjacent folded segments flow in opposite directions, validating the intended current-path extension mechanism. 
Fig.~3(a) plots the simulated co- and cross-polarized realized gain in both planes, showing a 2.55 dBi realized gain at the 5 GHz resonance. The resonance frequency of the proposed antenna can be tuned through the antenna width and vertical length, which are mainly controlled by \(L_{d}\) and \(L_{g}\), respectively. Figs.~3(b) and 3(c) plot the simulated reflection coefficient and maximum realized gain for varied \(L_{d}\) and \(L_{g}\), while all other dimensions are fixed. For \(L_{g}\) from 0.4 mm to 0.65 mm and \(L_{d}\) from 3.5 mm to 4.5 mm, the resonance frequency is tuned from 4.8 GHz to 5.3 GHz, while the maximum realized gain remains around 2.6 dBi.

\subsection{Four-element Dynamic Array}
Based on the theoretical analysis of the single planar meander-line monopole in Section~II-A, we develop a four-element broadside array on the same substrate, as shown in Fig.~4, with an overall size of $0.55\times1.73\times0.0083\lambda_{0}^{3}$. Four meander-line elements (Ant1, Ant2, Ant3, and Ant4) are linearly distributed along the \textit{x}-axis and oriented in the same direction. The element spacing \(d\) is one-half wavelength, which helps maintain low mutual coupling. A meander-line extension \(\Delta L_{d}\) for Ant1 is optimized to 1.1 mm to lower its resonance, as shown in Fig.~5. All elements are fed with in-phase currents and tuned at 5 GHz, yielding a maximum broadside realized gain of 7.4 dBi. For this design, the switching-state current vector \(\mathbf{i}^{(s)}\) can first be written as the column vector in \eqref{eq:state_current_general}, assuming a reference current amplitude \(I_0\). The manipulated current amplitudes are then described by introducing the excitation amplitude ratio \(\sqrt{\alpha}\).

\begin{equation}
	\mathbf{i}^{(s)} = I_0
	\begin{bmatrix}
		\gamma_1^{(s)} \\
		\gamma_2^{(s)} \\
		\gamma_3^{(s)} \\
		\gamma_4^{(s)}
	\end{bmatrix},
	\qquad
	\gamma_n^{(s)} \in \{1, \sqrt{\alpha}\},
	\label{eq:state_current_general}
\end{equation}
where $s \in \{A, B, C, D\}$ denotes the four switching states. The port current vector for each switching state can be expressed in a compact parametric form as
\begin{equation}
	\mathbf{i}^{(s)} = I_{0} \left( \mathbf{1} + \left( \sqrt{\alpha} - 1 \right) \mathbf{b}^{(s)} \right),
	\label{eq:state_current_selection}
\end{equation}
where $\mathbf{b}^{(s)}$ is a binary selection vector that identifies the dominant ports for state 
$s$. The selection vectors corresponding to the four switching states are compactly written as
\begin{equation}
\begin{aligned}
	\mathbf{b}^{(A)}&=\left[\begin{smallmatrix}0&1&1&0\end{smallmatrix}\right]^{T},&
	\mathbf{b}^{(B)}&=\left[\begin{smallmatrix}1&0&0&1\end{smallmatrix}\right]^{T},\\
	\mathbf{b}^{(C)}&=\left[\begin{smallmatrix}1&0&1&0\end{smallmatrix}\right]^{T},&
	\mathbf{b}^{(D)}&=\left[\begin{smallmatrix}0&1&0&1\end{smallmatrix}\right]^{T}.
\end{aligned}
	\label{eq:selection_vectors}
\end{equation}
where ``1'' and ``0'' indicate the strong and weak power states at each port, respectively. The normalized array factor $AF(\theta,\varphi)$ of the four-element array is expressed as
\begin{equation}
	AF(\theta,\varphi)
	= \sum_{n=1}^{4} \gamma_n e^{j k x_n \sin\theta \cos\varphi}.
\end{equation}
\begin{figure*}[t]
	\centering
\subfloat[]{\includegraphics[width=2.7in]{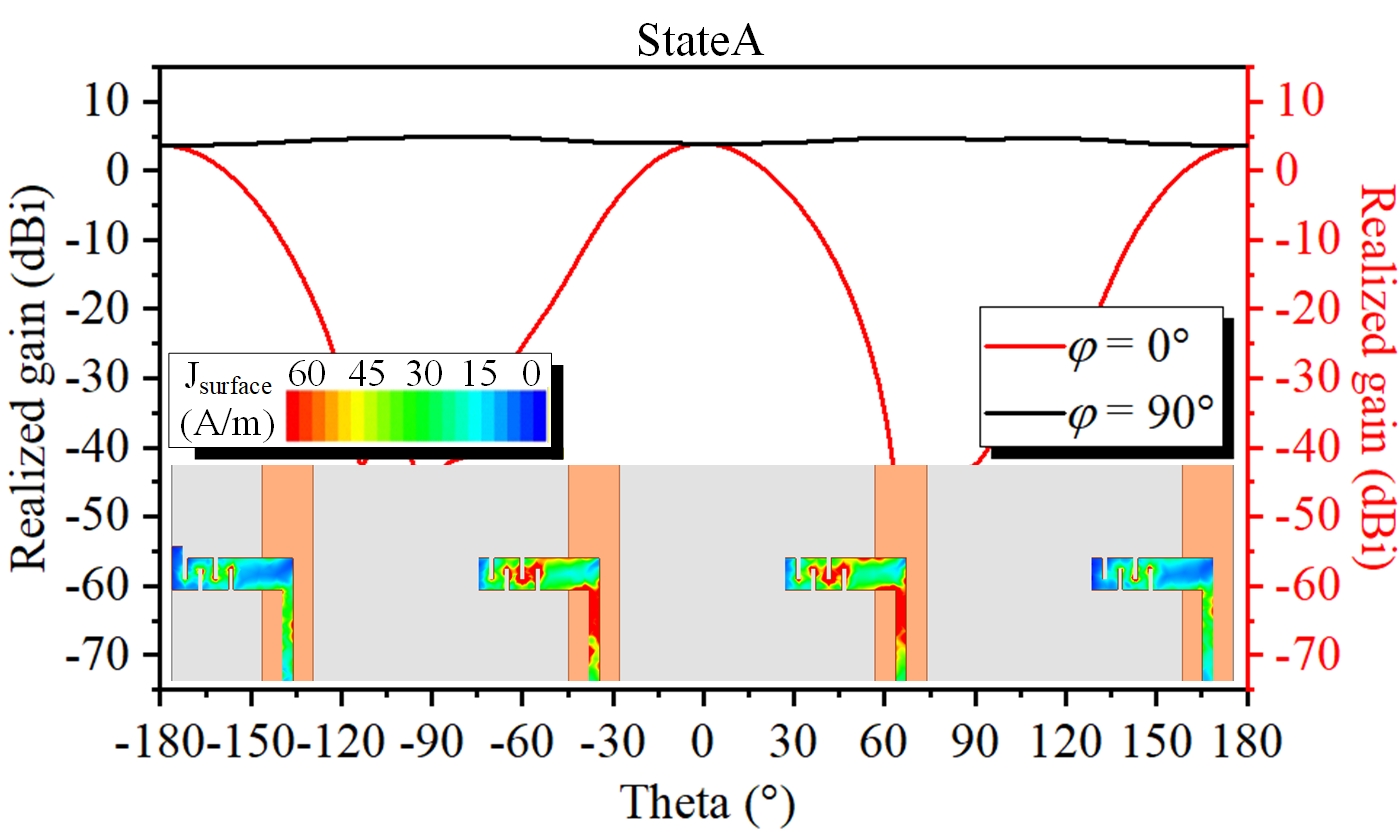}}\hspace{0.04\textwidth}
\subfloat[]{\includegraphics[width=2.7in]{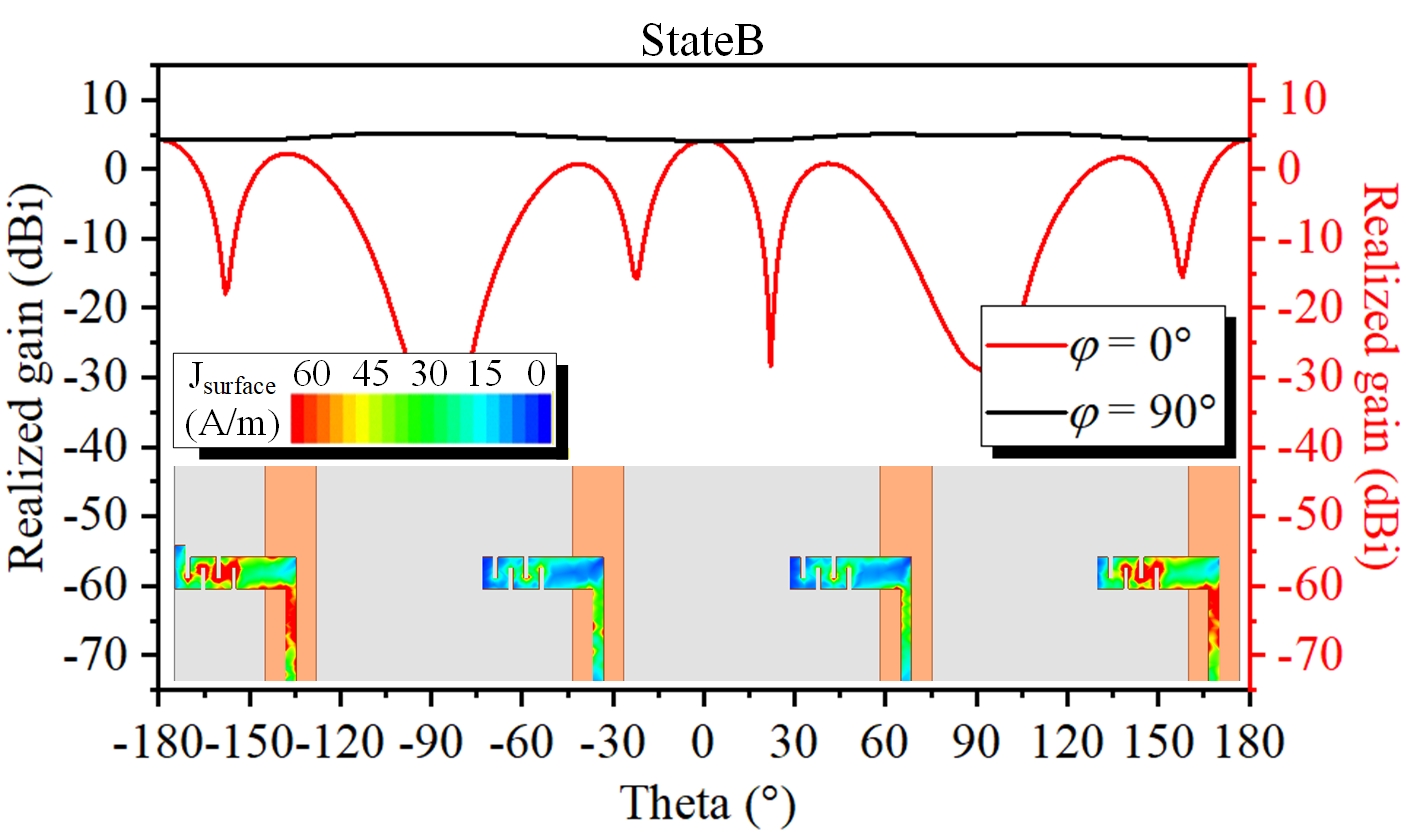}}\\
\subfloat[]{\includegraphics[width=2.7in]{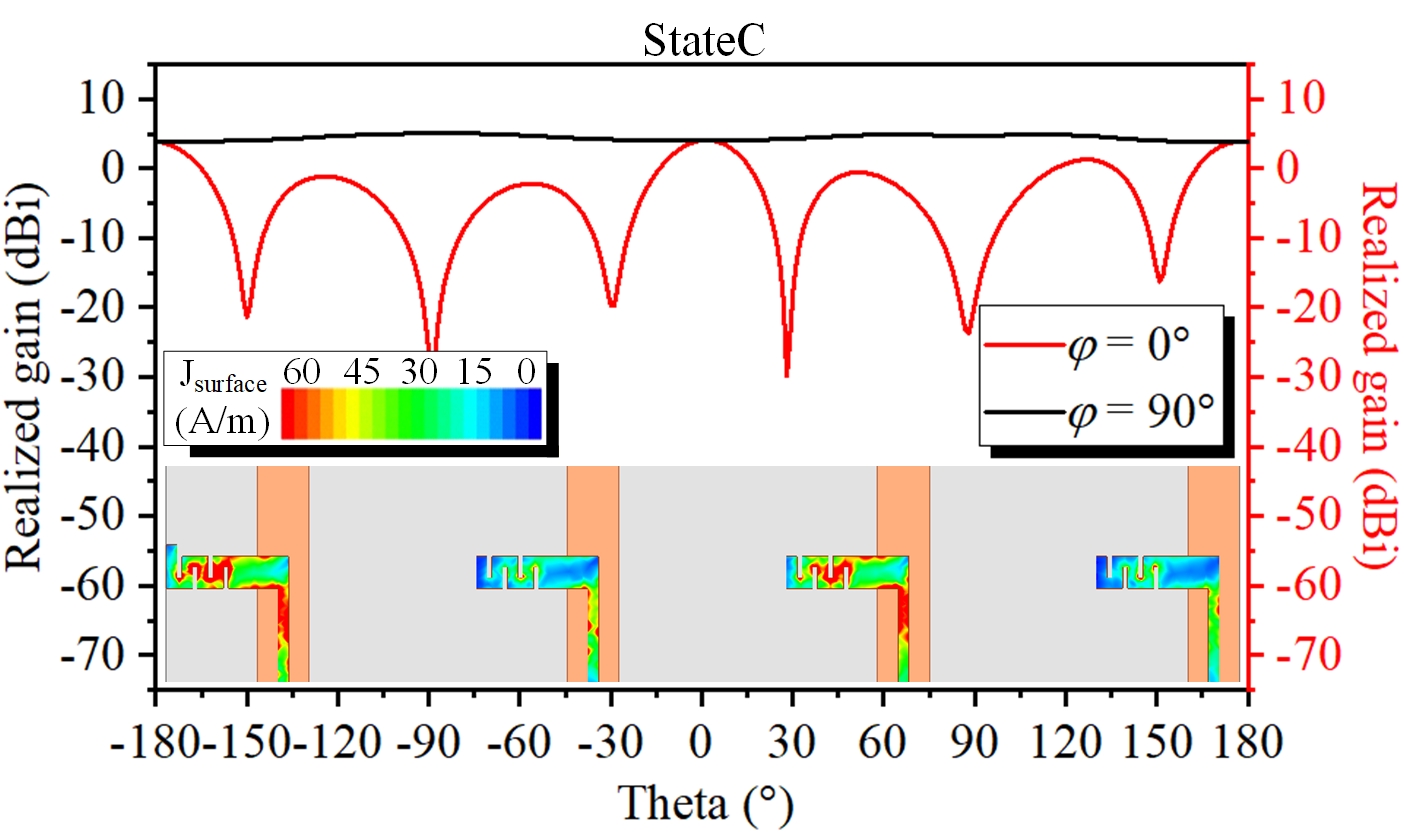}}\hspace{0.04\textwidth}
\subfloat[]{\includegraphics[width=2.7in]{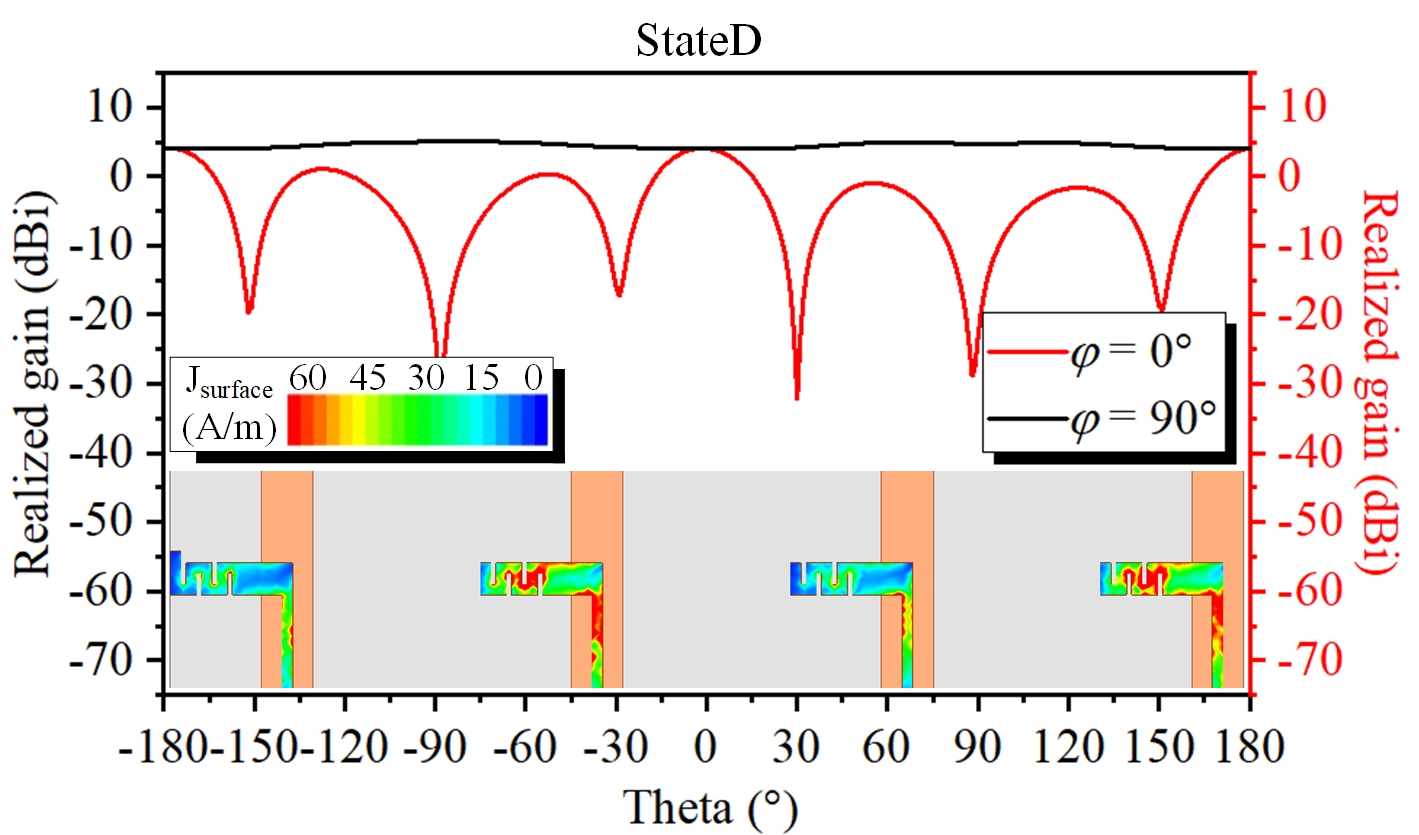}}
	\caption{Simulated surface current and corresponding realized gain radiation patterns of the proposed dynamic four-element array with a power ratio of $\alpha$ = 9 dB at 5 GHz for (a) State A, (b) State B, (c) State C, and (d) State D.}\label{fig:four_state_patterns}
\end{figure*}Therefore, the array factors for all states can be expressed as 
{\makeatletter
	\setlength{\@mathmargin}{0pt} 
	\begin{subequations}
		\begin{align}
			AF_{A}(\theta,\varphi)
			&=
			2
			\left[
			\sqrt{\alpha}\cos\!\left(\frac{3\psi}{2}\right)
			+
			\cos\!\left(\frac{\psi}{2}\right)
			\right], \label{eq:AF_A} \\
			AF_{B}(\theta,\varphi)
			&=
			2
			\left[
			\cos\!\left(\frac{3\psi}{2}\right)
			+
			\sqrt{\alpha}\cos\!\left(\frac{\psi}{2}\right)
			\right], \label{eq:AF_B} \\
			AF_{C}(\theta,\varphi)
			&=
			(1+\sqrt{\alpha})
			\left[
			\cos\!\left(\frac{3\psi}{2}\right)
			+
			\cos\!\left(\frac{\psi}{2}\right)
			\right] \nonumber \\
			&\quad
			+ j(1-\sqrt{\alpha})
			\left[
			\sin\!\left(\frac{3\psi}{2}\right)
			-
			\sin\!\left(\frac{\psi}{2}\right)
			\right], \label{eq:AF_C} \\
			AF_{D}(\theta,\varphi)
			&=
			(1+\sqrt{\alpha})
			\left[
			\cos\!\left(\frac{3\psi}{2}\right)
			+
			\cos\!\left(\frac{\psi}{2}\right)
			\right] \nonumber \\
			&\quad
			+ j(\sqrt{\alpha}-1)
			\left[
			\sin\!\left(\frac{3\psi}{2}\right)
			-
			\sin\!\left(\frac{\psi}{2}\right)
			\right], \label{eq:AF_D}
		\end{align}
	\end{subequations}
	\makeatother}where $\psi = k d \sin\theta \cos\varphi$ is the spatial phase term. When switching between states A $\&$ B, states C $\&$ D, and all states switching, the differential terms $AF_{\Delta_{AB}}$, $AF_{\Delta_{CD}}$, and $AF_{\Delta_{ABCD}}$ of the array factor can be calculated as

\begin{subequations}
	\begin{align}
		AF_{\Delta_{AB}} 
		&= \left( \sqrt{\alpha} - 1 \right)
		\left[
		\cos\left( \frac{3\psi}{2} \right)
		-
		\cos\left( \frac{\psi}{2} \right)
		\right],
		\label{eq:AF_delta_AB}
		\\
		AF_{\Delta_{CD}}
		&= j\left(1 - \sqrt{\alpha}\right)
		\left[
		\sin\left(\frac{3\psi}{2}\right)
		-
		\sin\left(\frac{\psi}{2}\right)
		\right],
		\label{eq:AF_delta_CD}
		\\
		AF_{\Delta_{ALL}}
		&= (\sqrt{\alpha}-1)
		\left[
		\cos\frac{3\psi}{2}
		-
		\cos\frac{\psi}{2}
		\right]
		\nonumber \\
		&\quad + j(1-\sqrt{\alpha})
		\left[
		\sin\frac{3\psi}{2}
		-
		\sin\frac{\psi}{2}
		\right].
		\label{eq:AF_delta_ALL}
	\end{align}
\end{subequations}
These expressions reveal that the A-B switching pair produces a purely real differential field, corresponding primarily to amplitude perturbations, whereas the C-D switching pair generates an imaginary differential field, resulting in a quadrature component that manifests as phase or complex-gain modulation. In both cases, the switching-induced distortion vanishes when $\alpha=1$, confirming that directional modulation arises exclusively from the intentional power imbalance between the excitation states.
When switching among all four states (A-B-C-D) with equal time weighting, the combined differential component can be decomposed into the A-B and C-D differential components, i.e., $AF_{\Delta,ALL} = AF_{\Delta,AB} + AF_{\Delta,CD}$.
Unlike the pairwise switching cases, the four-state operation therefore introduces both real and imaginary dynamic-field perturbations over one switching period, corresponding to combined magnitude and phase modulation in the spatial domain.
For the transmitted power of the dynamic array, the time-averaged radiated power under equal state weighting contains contributions from both the average field and the switching-induced dynamic field variation:
\begin{subequations}
	\begin{align}
		\bar{P}_{AB}(\theta,\varphi) 
		&\propto 
		\left|AF_{\mathrm{avg},AB}(\theta,\varphi)\right|^2 
		+
		\left|AF_{\Delta,AB}(\theta,\varphi)\right|^2, 
		\label{eq:P_AB} 
		\\
		\bar{P}_{CD}(\theta,\varphi) 
		&\propto 
		\left|AF_{\mathrm{avg},CD}(\theta,\varphi)\right|^2 
		+
		\left|AF_{\Delta,CD}(\theta,\varphi)\right|^2,
		\label{eq:P_CD}
		\\
		\bar{P}_{ALL}(\theta,\varphi) 
		&\propto 
		\left|AF_{\mathrm{avg},ALL}(\theta,\varphi)\right|^2 
		+
		\left|AF_{\Delta,ALL}(\theta,\varphi)\right|^2,
		\label{eq:P_ALL}
	\end{align}
\end{subequations}
where $AF_{\mathrm{avg},AB}=(AF_A+AF_B)/2$ and $AF_{\mathrm{avg},CD}=(AF_C+AF_D)/2$, respectively.

The angle-dependent effective SNR can be expressed as
\begin{equation}
	\mathrm{SNR}_{\mathrm{eff}}(\theta,\varphi)
	=
	\frac{|AF_{\mathrm{avg}}|^2 P_s}{|AF_\Delta|^2 P_s + P_n},
\end{equation}
which directly links the array dynamics to demodulation performance.
Here, $P_s$ and $P_n$ denote the average symbol power and noise power, respectively, representing the ratio between the power of the average array factor and the combined power of the switching-induced differential component and additive noise.
Under high-SNR conditions, the switching-induced dynamics dominate the performance, such that as $|AF_\Delta|$ increases, $\mathrm{SNR}_{\mathrm{eff}}$ decreases, resulting in a higher BER.

Fig.~6 plots the simulated surface current on the radiators and the corresponding realized gain radiation patterns for the four static states with an identical power ratio of $\alpha=9$ dB. Stronger current density is obtained on the dominant radiators in each state. The realized gain patterns show that, for all states in the E-plane ($\varphi=0^\circ$), the main radiation direction remains near broadside and the beam is approximately symmetric. For the A--B switching pair, a large differential gain appears at off-broadside angles because the phase center remains nearly fixed while the excitation magnitudes are exchanged between symmetric radiators. For the C--D switching pair, the gain patterns are nearly identical, while the phase pattern changes sign because the amplitude distribution remains approximately unchanged.

\begin{figure}[t]
	\centering
\includegraphics[width=3in]{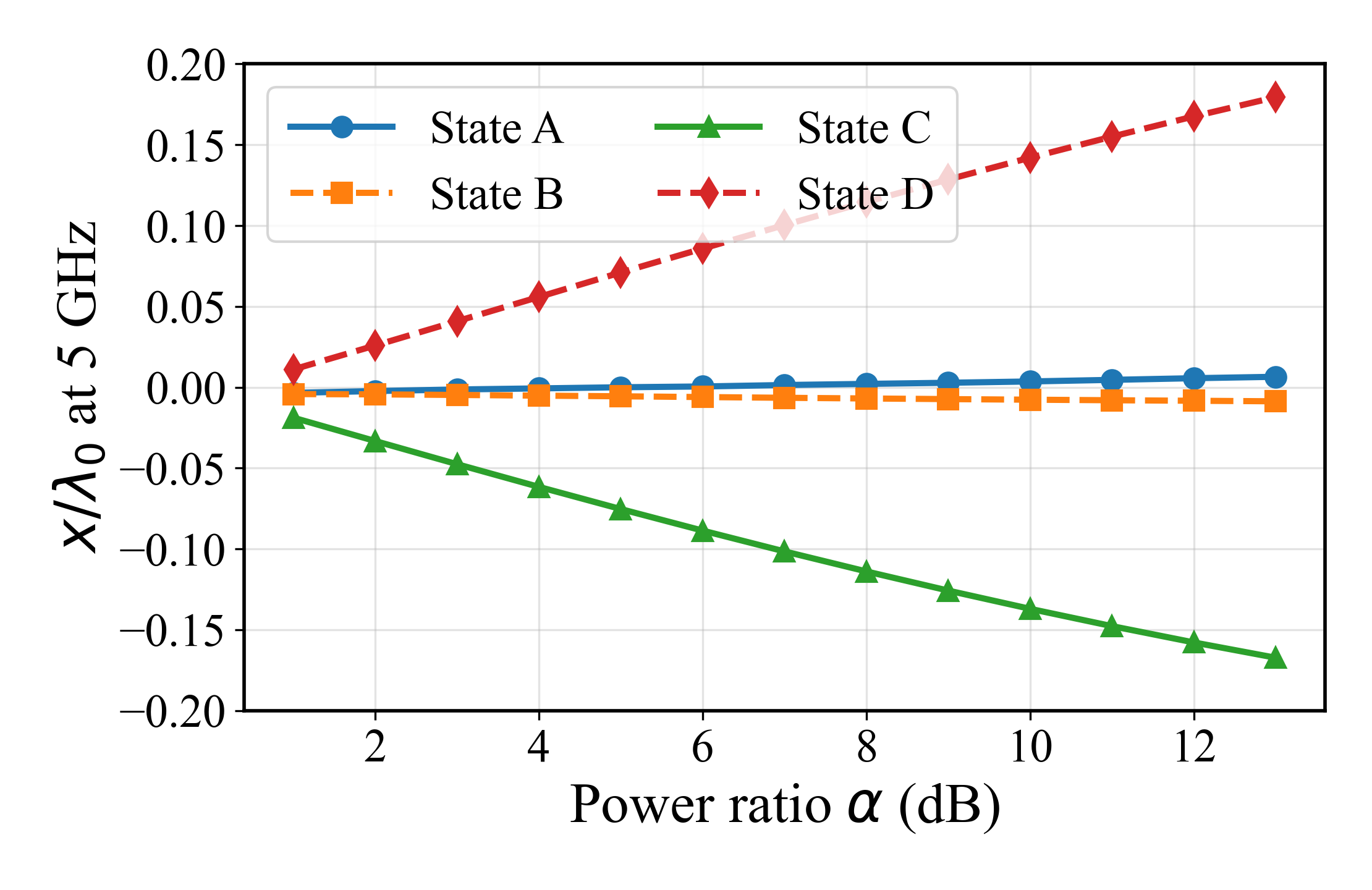}
	\caption{Simulated phase-center displacement along the $x$ direction as a function of power ratio $\alpha$ for all static states.}
	\label{fig:phase_center_power_ratio}
\end{figure}

The phase-center displacement in Fig.~\ref{fig:phase_center_power_ratio} was extracted from the simulated phase patterns using the mean electrical phase center (MEPC) procedure \cite{kumar2013improved,10286341}. For each value of $\alpha$, the radiated phase was first unwrapped within a broadside angular window of $-10^\circ \leq \theta \leq 10^\circ$, where the positive $z$ axis defines broadside. Adjacent-angle phase differences were then used to form local phase-center estimates, and the resulting cluster was averaged to obtain the MEPC along the broadside. The plotted quantity is the corresponding lateral displacement along $x$, normalized to the free-space wavelength at 5~GHz, which captures the phase-center motion produced by changing the relative excitation amplitudes of the four static states. It can be seen that due to the differential power, the phase-center of states C and D moves away from the broadside, while for states A and B, the phase-center is almost maintained at the center. C--D switching mainly reproduces lateral phase-center motion, and A--B switching reveals a different mechanism whereby the fixed phase-center creates a dynamic element spacing effect.

\begin{figure*}[t]
	\centering
\subfloat[]{\includegraphics[width=3.2in]{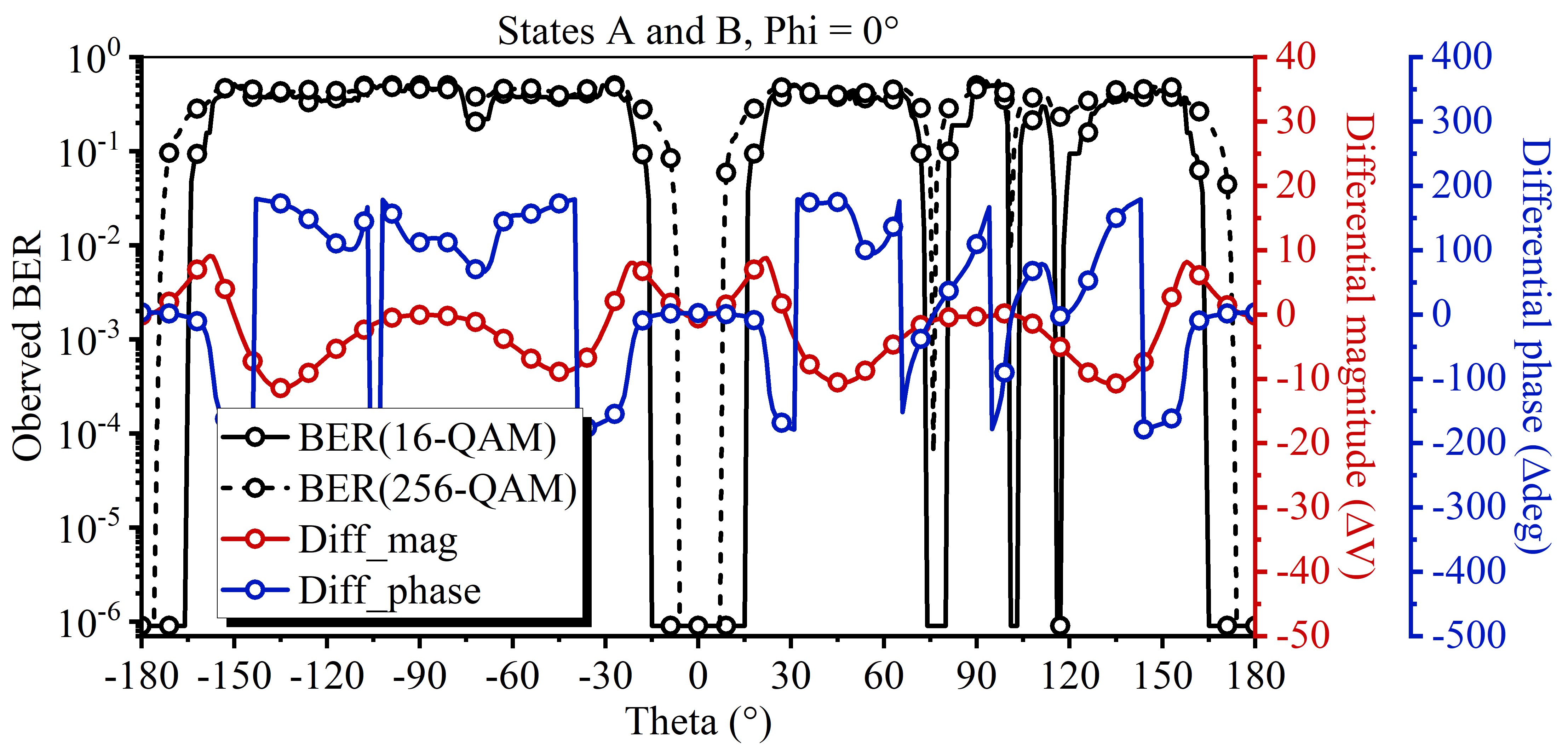}}\hfil
\subfloat[]{\includegraphics[width=3.2in]{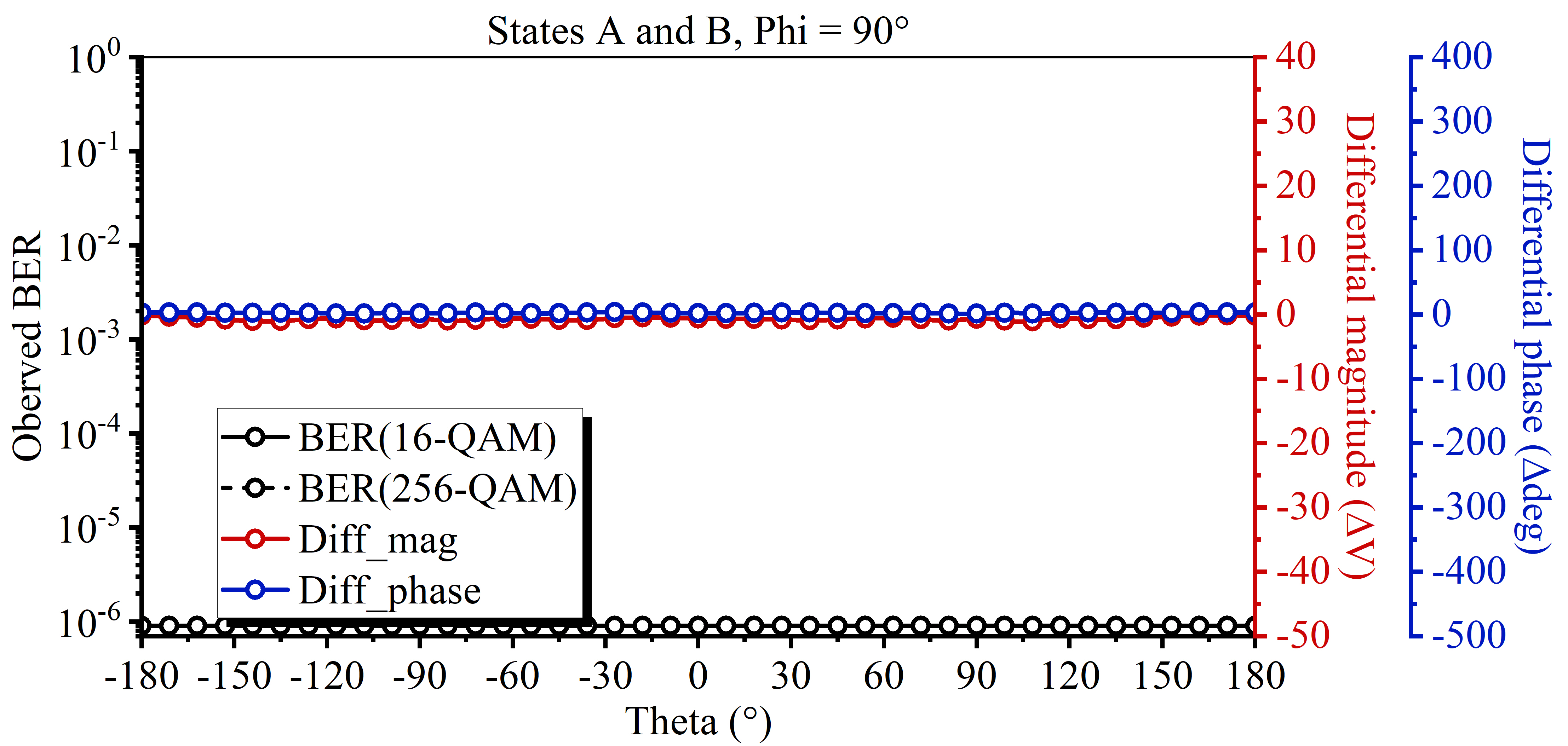}}\hfil\\
\subfloat[]{\includegraphics[width=3.2in]{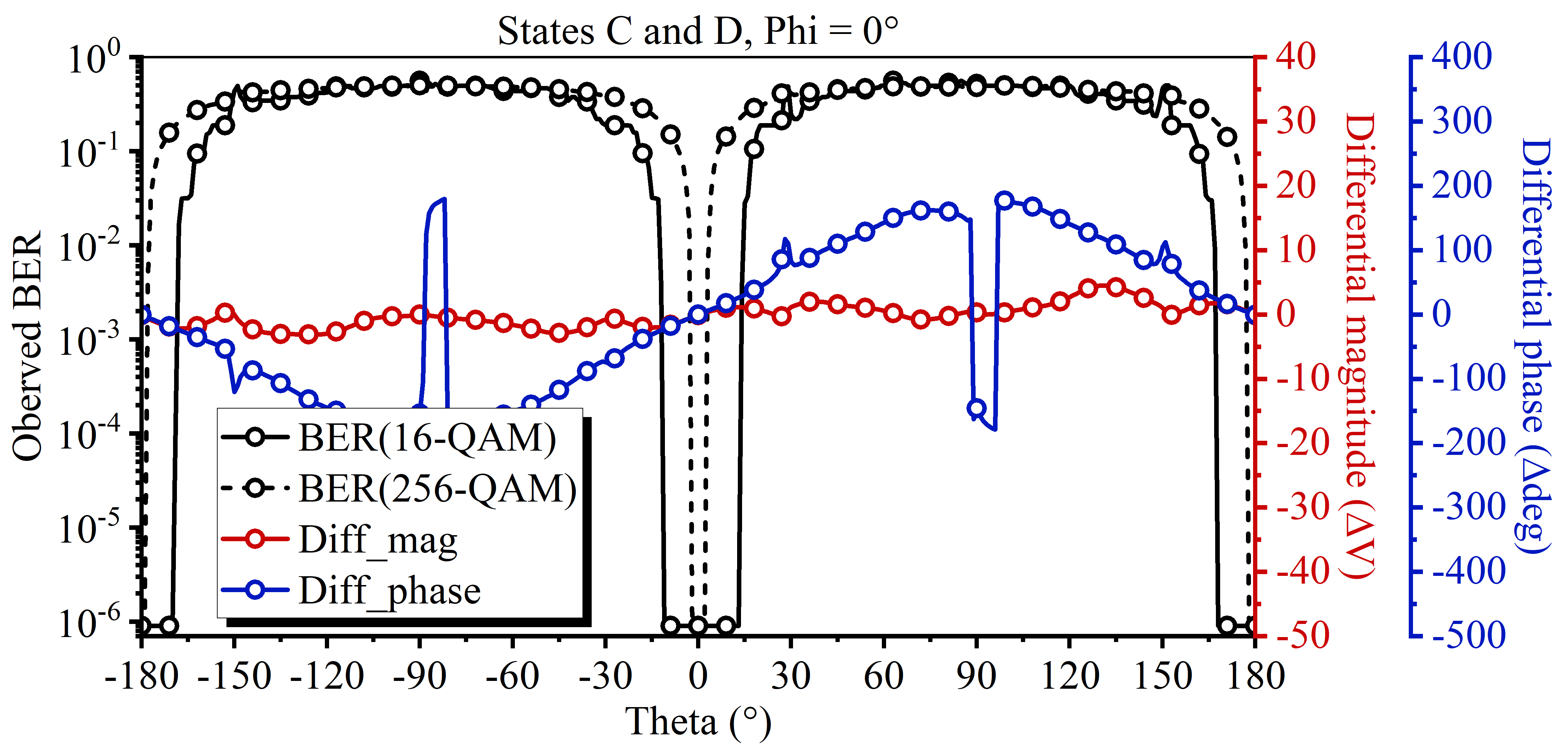}}\hfil
\subfloat[]{\includegraphics[width=3.2in]{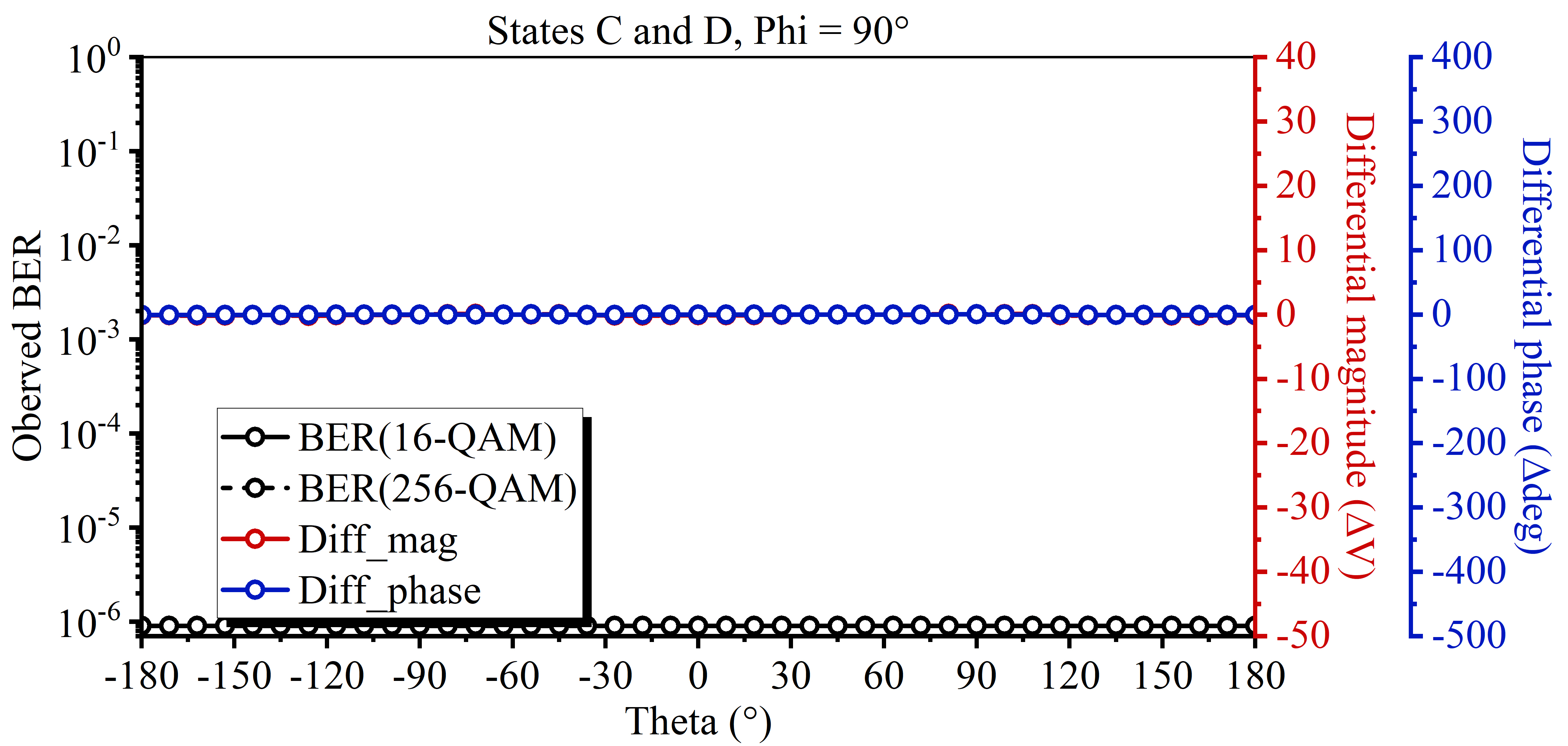}}\hfil\\
\subfloat[]{\includegraphics[width=3.2in]{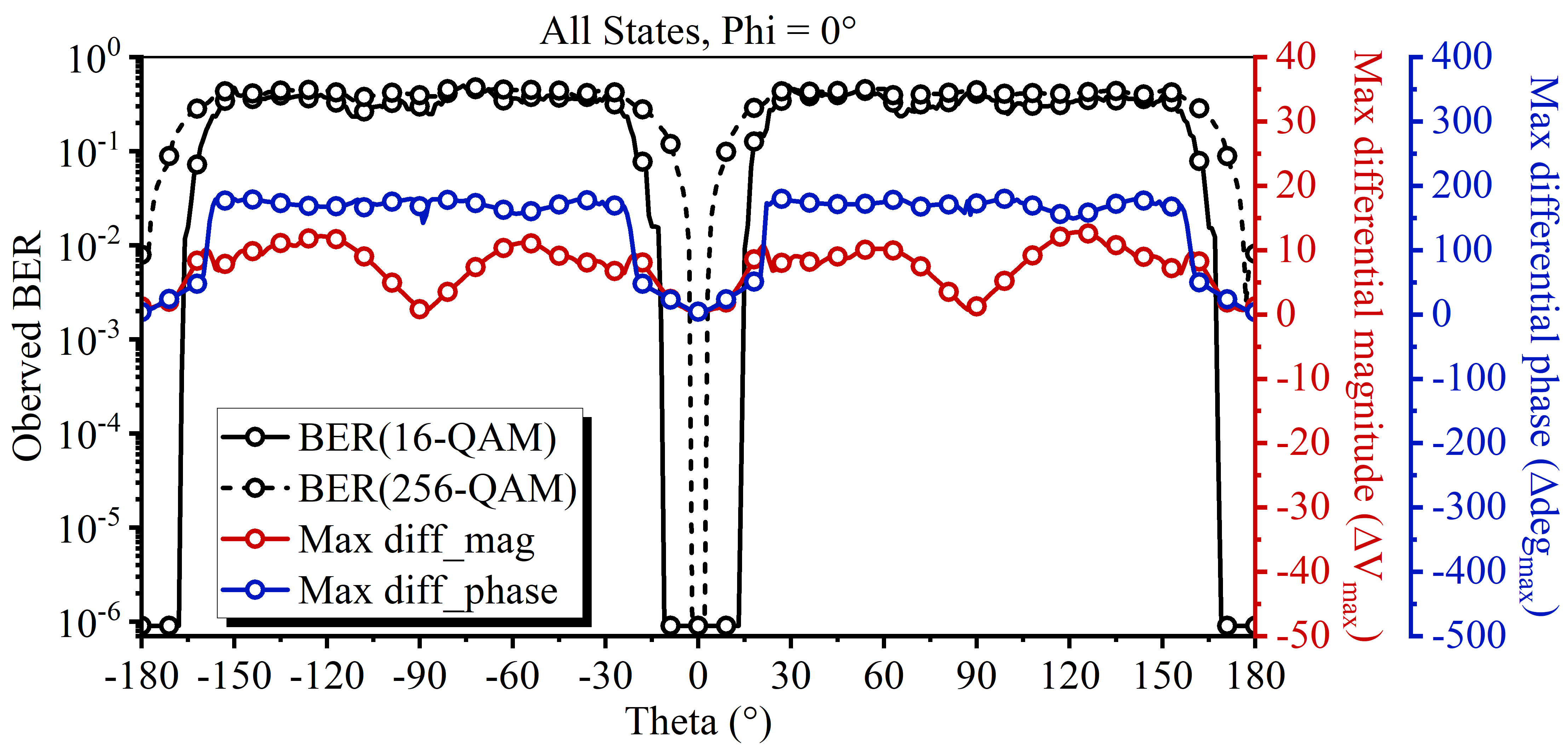}}\hfil
\subfloat[]{\includegraphics[width=3.2in]{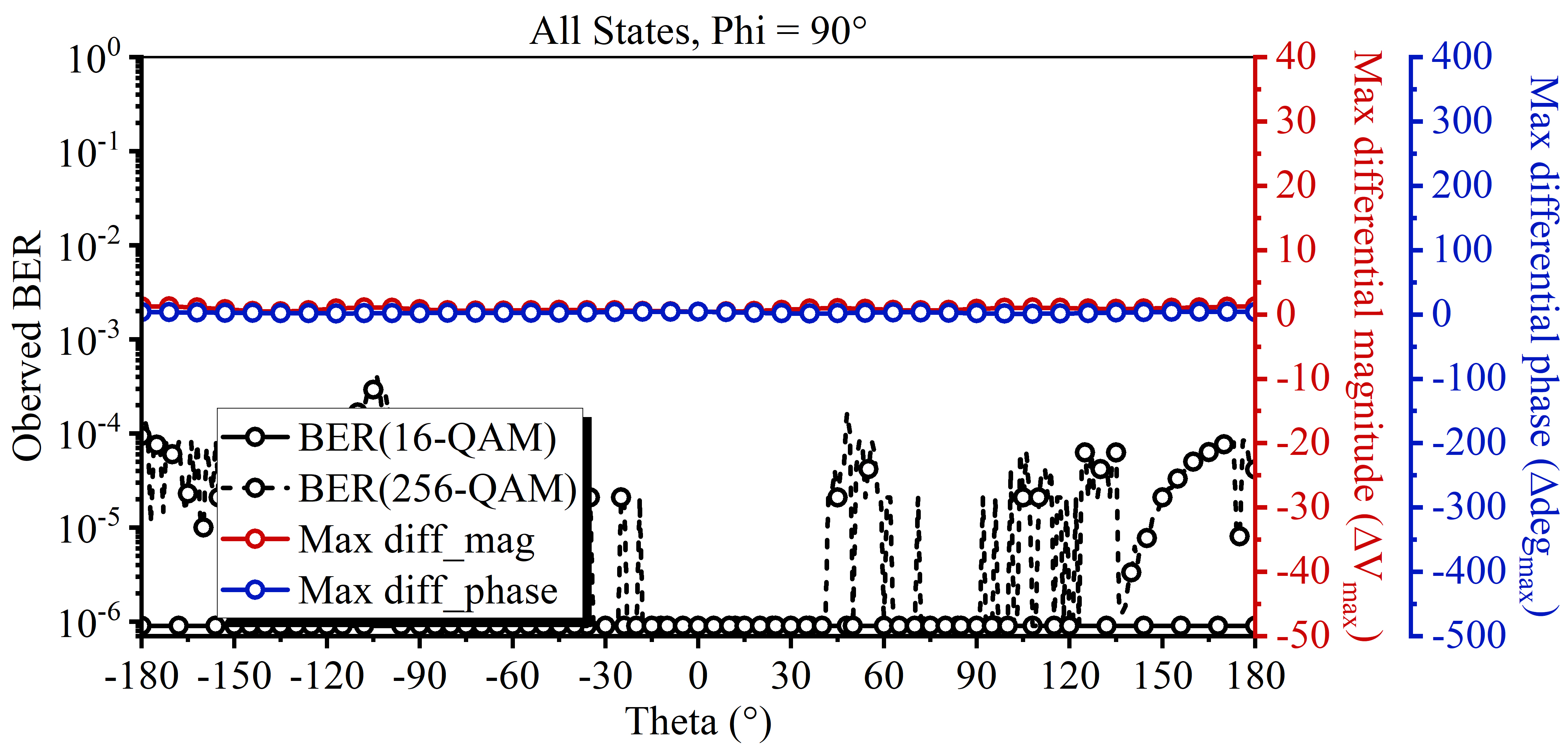}}\hfil
	\caption{Simulated communication performance of the proposed dynamic array at 5~GHz with a power ratio $\alpha = 9$~dB. The observed BER and the corresponding differential magnitude and phase versus $\theta$ are presented for (a) and (b) switching between states A and B in the E- and H-planes, respectively; (c) and (d) switching between states C and D; and (e) and (f) full four-state (ABCD) switching in the E- and H-planes, respectively. For the two-state cases, $\Delta \mathrm{V}$ and $\Delta \mathrm{deg}$ denote the pairwise magnitude and wrapped phase differences between the two switching states. For the four-state cases, $\Delta \mathrm{V}_{\max}$ and $\Delta \mathrm{deg}_{\max}$ denote the maximum pairwise magnitude and wrapped phase differences at each observation angle among all six state combinations: A--B, A--C, A--D, B--C, B--D, and C--D.}
	
\end{figure*}
\begin{figure}[t]
	\centering
\includegraphics[width=3.1in]{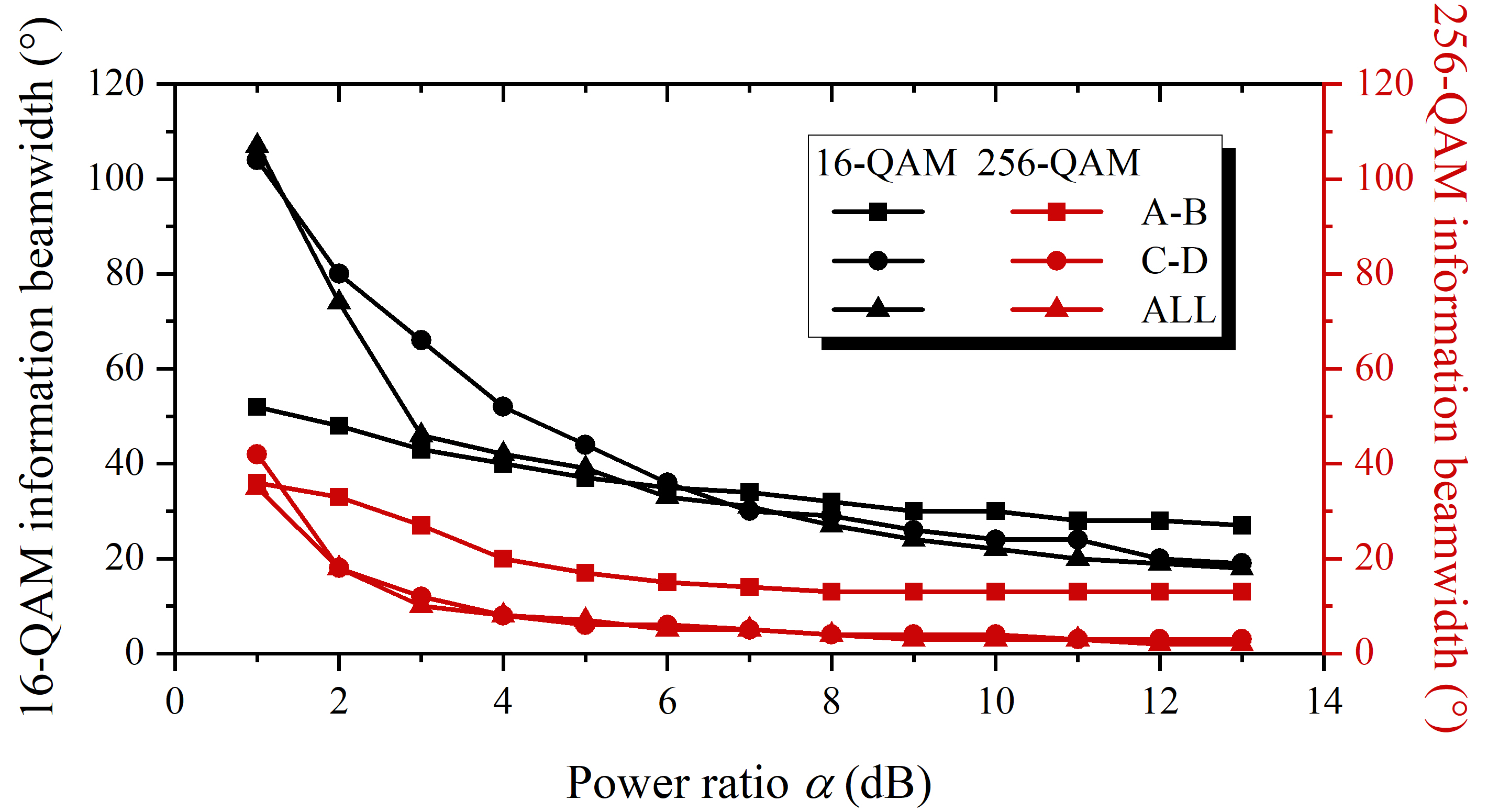}%
	\hfil\hfil
	\caption{Simulated E-plane information beamwidth of the proposed dynamic array using 16- and 256-QAM signals for varied power ratio $\alpha$ under different switching cases.}
\end{figure}

\section{Analysis of the Information Beamwidth}

In this section, we study the communication performance of the designed dynamic omnidirectional meander antenna in MATLAB using the communication channel model in \cite{10286341}. A 48-kb pseudorandom bit sequence was modulated onto 16- and 256-QAM signals using Gray coding, and the simulated amplitude and phase patterns exported from HFSS were then applied to the channel model. An SNR of 40 dB was used to ensure that the BER is dominated by directional modulation rather than low received SNR. We also evaluate the information beamwidth (IB), defined as the angular span over which BER $\leq 10^{-3}$, to indicate the region where the transmitted information can be recovered. Fig.~8 plots the simulated BER with the corresponding differential magnitude and phase for two switching cases in both the E- and H-planes using 16- and 256-QAM signals with amplitude ratios of $\pm$9 dB. Higher-order QAM produces a narrower IB. The angular dependence of the BER can be directly interpreted using the average--differential array factor decomposition derived in Section~II. Under dynamic switching, the time-averaged radiated power is given by $\bar{P}(\theta,\varphi) \propto |AF_{\mathrm{avg}}(\theta,\varphi)|^2 + |AF_{\Delta}(\theta,\varphi)|^2$, where the average term governs the intended radiation behavior, while the differential term captures the switching-induced amplitude and phase perturbations. For both switching cases, the BER is dominated by the relative contribution of the differential term in angular regions where switching-induced distortion outweighs the average radiation component.

For a fixed power ratio of $\alpha=9$~dB, different switching configurations supported by the same antenna array lead to different angular distributions of the differential array factor $AF_{\Delta}(\theta,\varphi)$. Although the functional form of $\bar{P}(\theta,\varphi)$ remains unchanged, the spatial extent and magnitude of the differential term vary with the selected switching pair, producing distinct angular patterns of waveform distortion and BER.

Specifically, the A-B switching configuration generates a comparatively larger differential magnitude at off-broadside angles, resulting in elevated values of $|AF_{\Delta}(\theta,\varphi)|$ over a wider angular range. According to the power decomposition in $\bar{P}(\theta,\varphi)$, this increases the relative weight of the differential term away from broadside, leading to stronger waveform corruption and higher BER in off-axis directions. It is also worth noting that around the off-broadside angles, A-B switching also introduces increased differential phase in the side lobes. This is because the phase centers of the two antenna elements on the sides are shifted and consequently, A-B switching introduces dynamic motion of the apparent element spacing, enhancing physical-layer security against receivers located outside the intended broadside region by deliberately amplifying switching-induced distortion at those angles.

In contrast, the C-D switching configuration produces a more spatially localized differential contribution, such that $|AF_{\mathrm{avg}}(\theta,\varphi)|^2$ dominates only within a narrower angular region. This confines low-BER communication to a tighter information beamwidth while rapidly increasing BER outside this region. Importantly, both switching modes operate under the same average--differential framework; however, they emphasize different security characteristics: A-B switching enhances robustness against off-broadside receivers through stronger differential magnitude, whereas C-D switching achieves improved angular selectivity by narrowing the region where the average term remains dominant. In Fig.~9, we calculate the achievable IB for the two switching cases using 16- and 256-QAM over varied power ratios, providing design guidance for selecting the amplitude dynamics.

\begin{figure*}[t]
	\centering
	\subfloat[]{%
\includegraphics[width=5.8in]{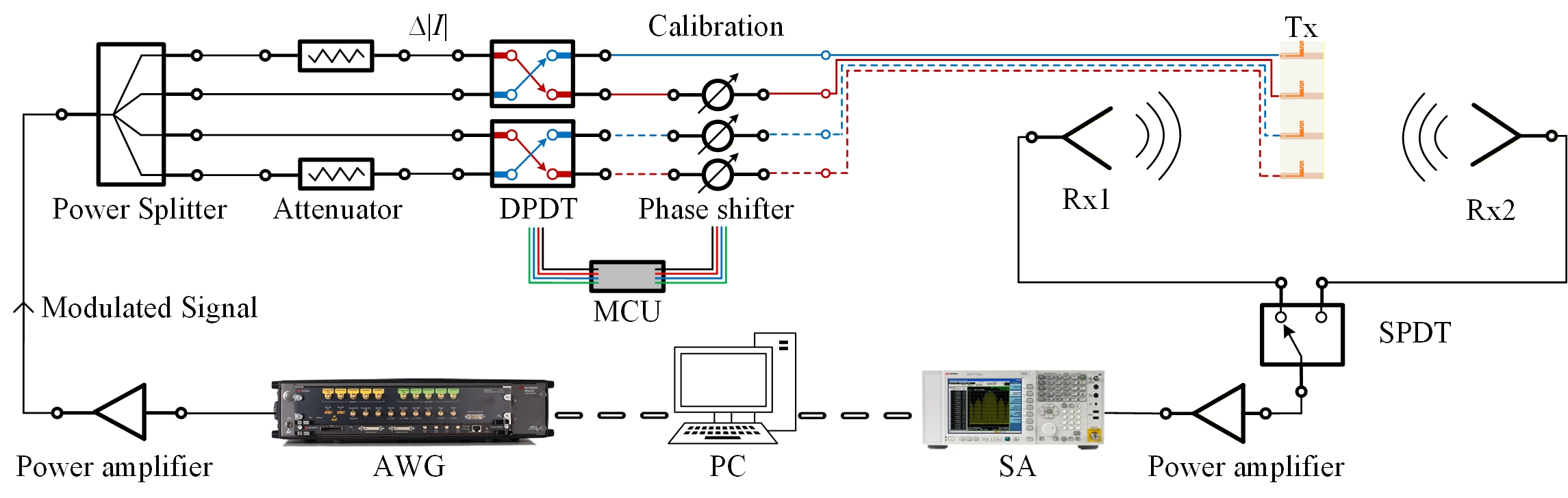}
	}\hfil
	\subfloat[]{%
\includegraphics[width=3.5in]{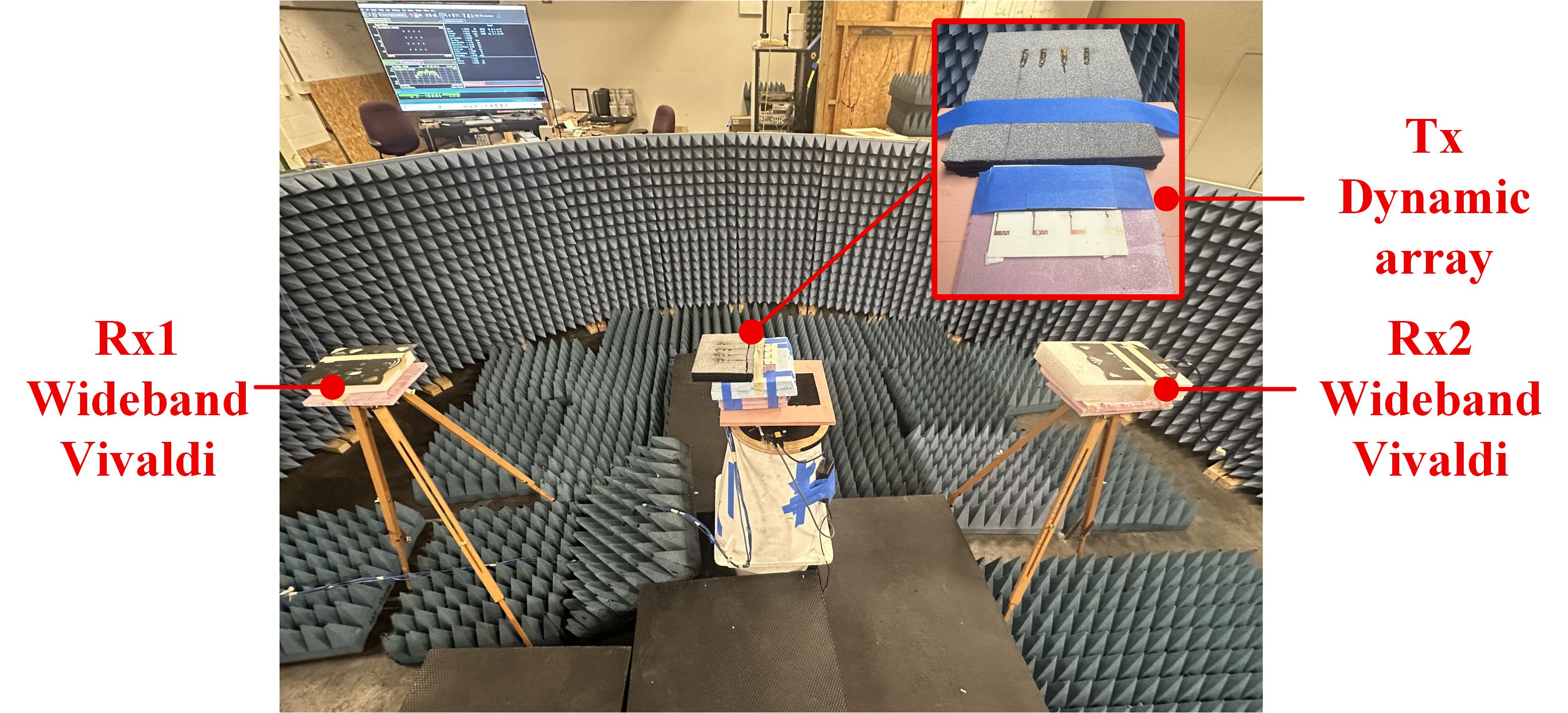}
	}\hfil
	\subfloat[]{%
\includegraphics[width=3.3in]{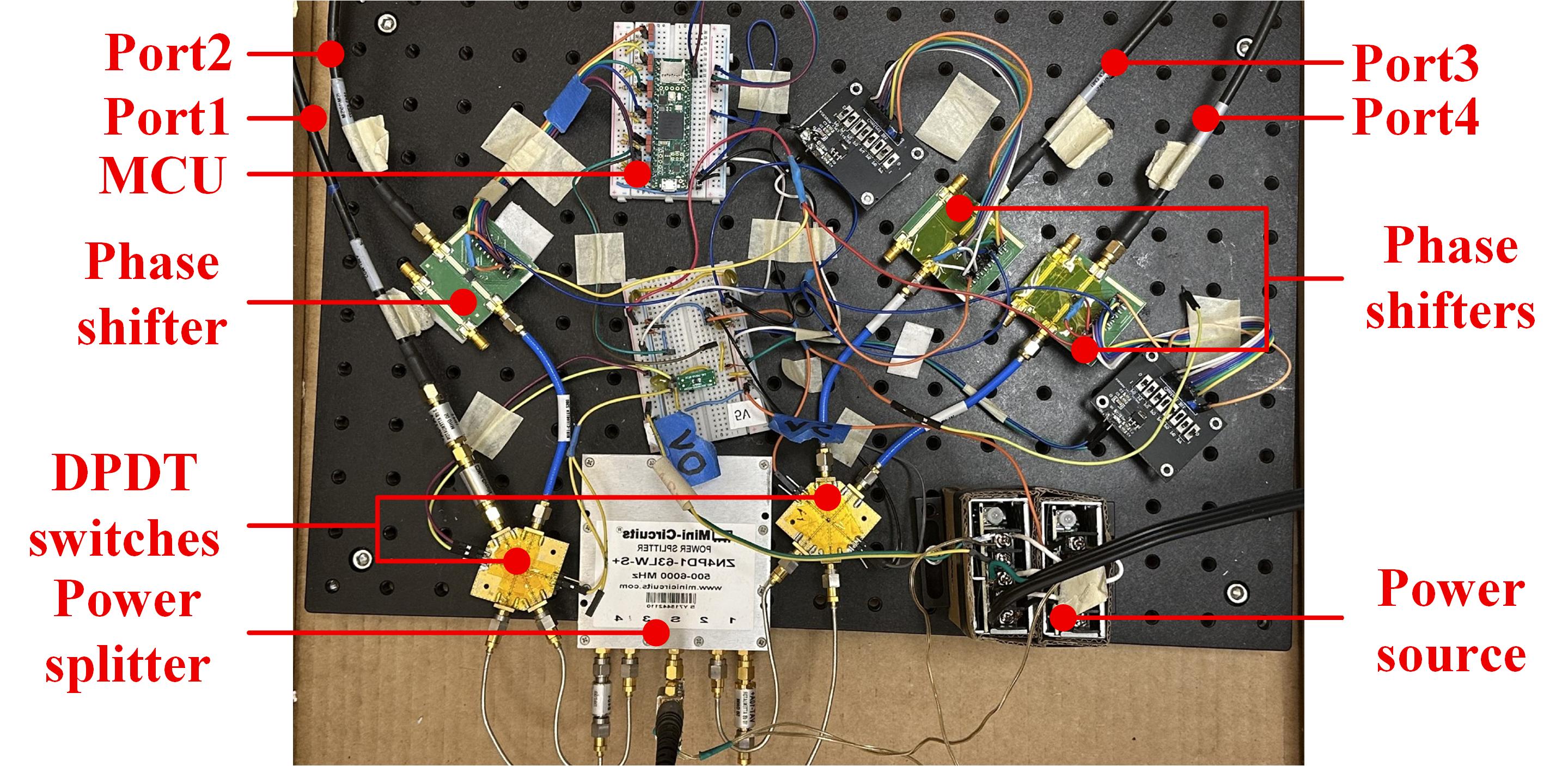}
	}
	
	\caption{Measurement setup for evaluating the communication performance of the proposed dynamic array: (a) block diagram of the measurement system, (b) experimental measurement setup, and (c) switching system configuration.}
	\label{fig:measurement_setup}
\end{figure*}
In addition to the E-plane results, the corresponding H-plane performance further validates the omnidirectional nature of the proposed antenna. As shown in Figs.~8(b) and~8(d), both the observed BER and the differential magnitude and phase remain nearly invariant with respect to $\theta$ for $\varphi=90^\circ$, regardless of the selected switching configuration. This behavior indicates that the average array factor dominates uniformly in the H-plane, while the differential contribution remains spatially uniform and does not introduce angularly dependent waveform distortion. Consequently, the communication performance is preserved over all azimuth angles, confirming that the proposed dynamic array maintains omnidirectional coverage in the H-plane while selectively enforcing security through switching-induced distortion in the E-plane. This observation is consistent with the average--differential power formulation, since a nearly constant $AF_{\Delta}(\theta,\varphi)$ in the H-plane results in an angle-independent contribution to $\bar{P}(\theta,\varphi)$, thereby preventing directional degradation of BER and preserving omnidirectional radiation characteristics.
\begin{figure}[t]
	\centering
\includegraphics[width=2.2in]{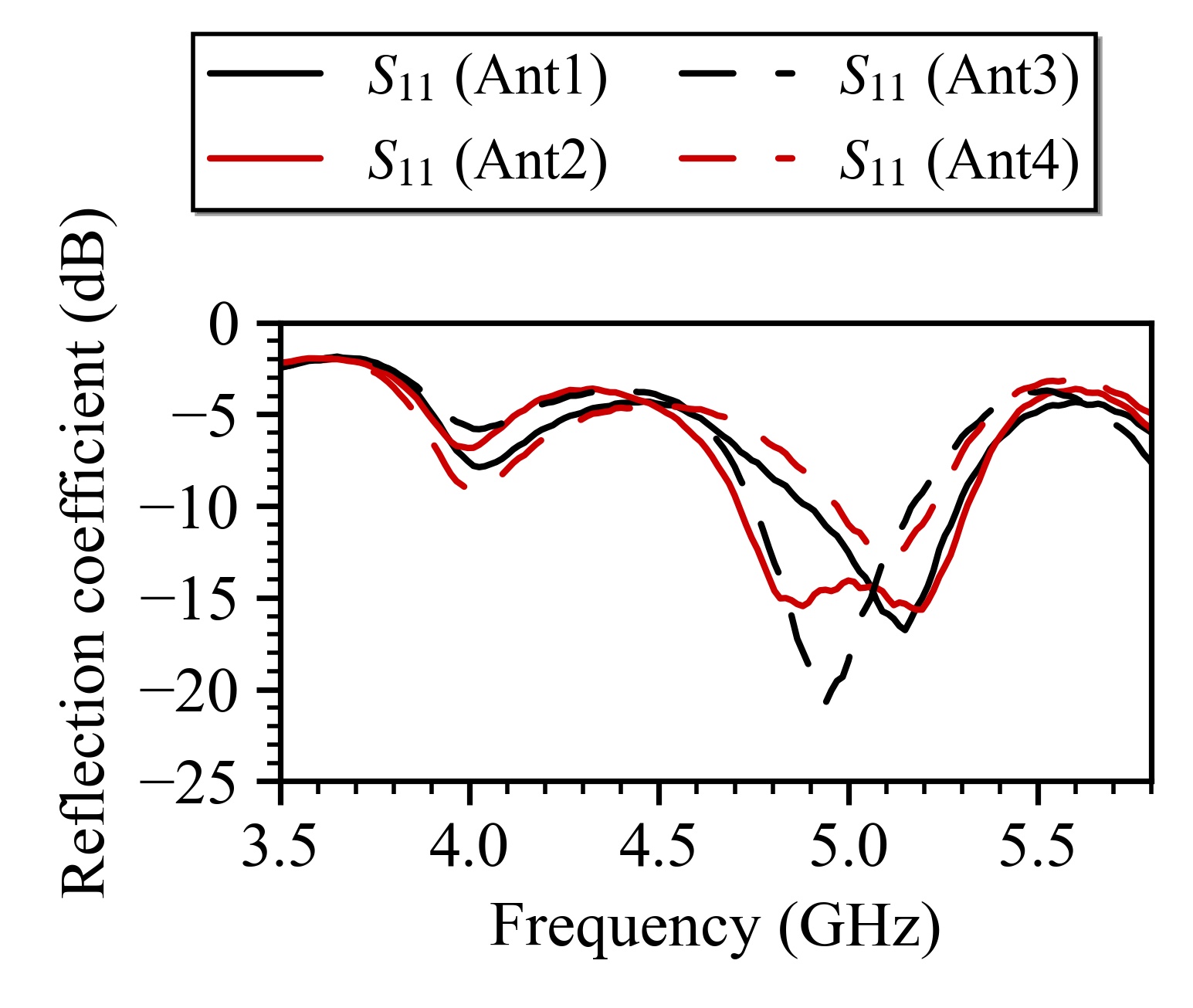}
	\caption{Measured reflection coefficient of the four-element array.}
	\label{fig:measured_s11}
\end{figure}

When switching among all four states, the dynamic array simultaneously incorporates both the real differential component associated with A-B switching and the quadrature component associated with C-D switching. Consequently, the resulting differential array factor contains combined magnitude and phase perturbations, leading to enhanced waveform distortion outside the intended broadside region. 

As shown in Figs. 8(e) and (f), the full four-state switching produces the most confined information beam in the E-plane, where reliable demodulation is restricted to a narrow angular sector around broadside. This behavior can be interpreted within the average differential decomposition framework: the simultaneous presence of real and imaginary differential components increases the relative contribution of $|AF_{\Delta}(\theta,\varphi)|$ away from broadside.
In contrast, the H-plane response remains largely invariant due to the preserved structural symmetry of the array, resulting in quasi-uniform communication performance over all azimuth angles. These results confirm that full four-state switching enhances planar selectivity by jointly exploiting magnitude and phase modulation while maintaining omnidirectional coverage in the orthogonal plane.

\begin{figure*}[t]
\subfloat[]{\includegraphics[width=0.49\textwidth]{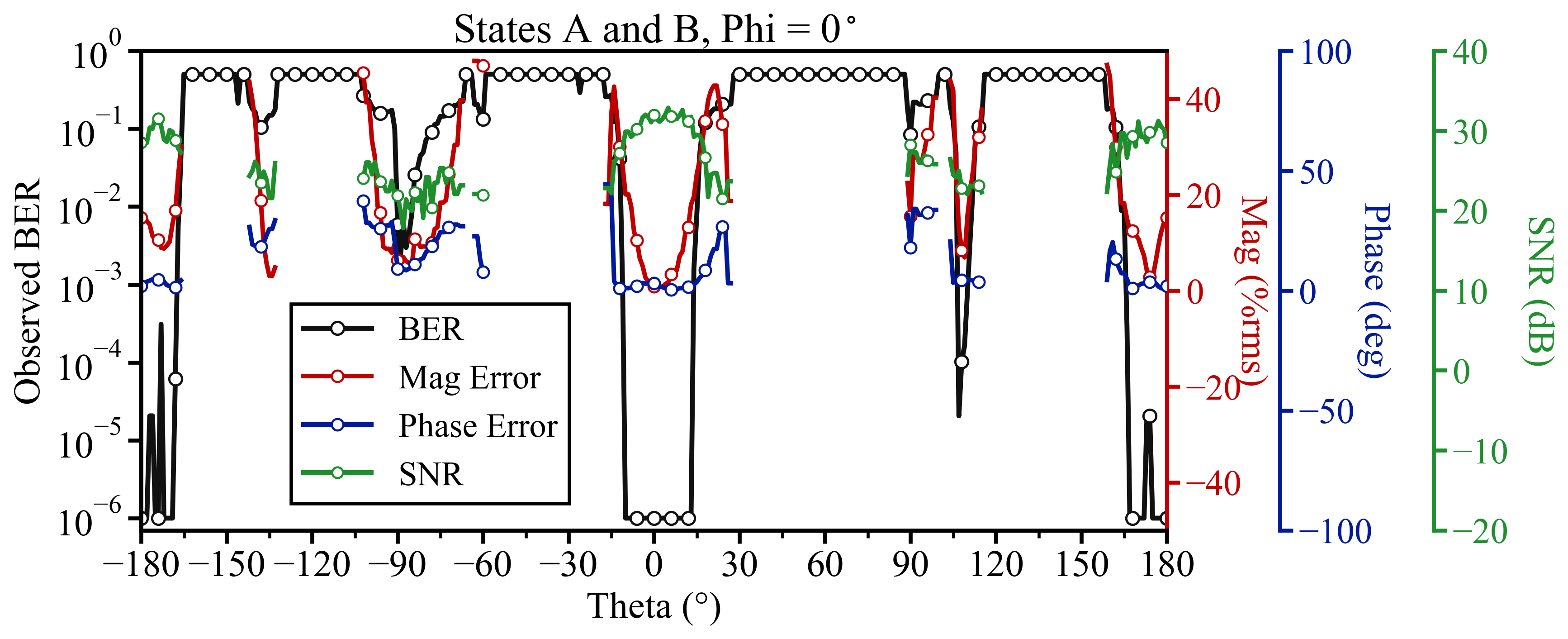}}\hfil
\subfloat[]{\includegraphics[width=0.49\textwidth]{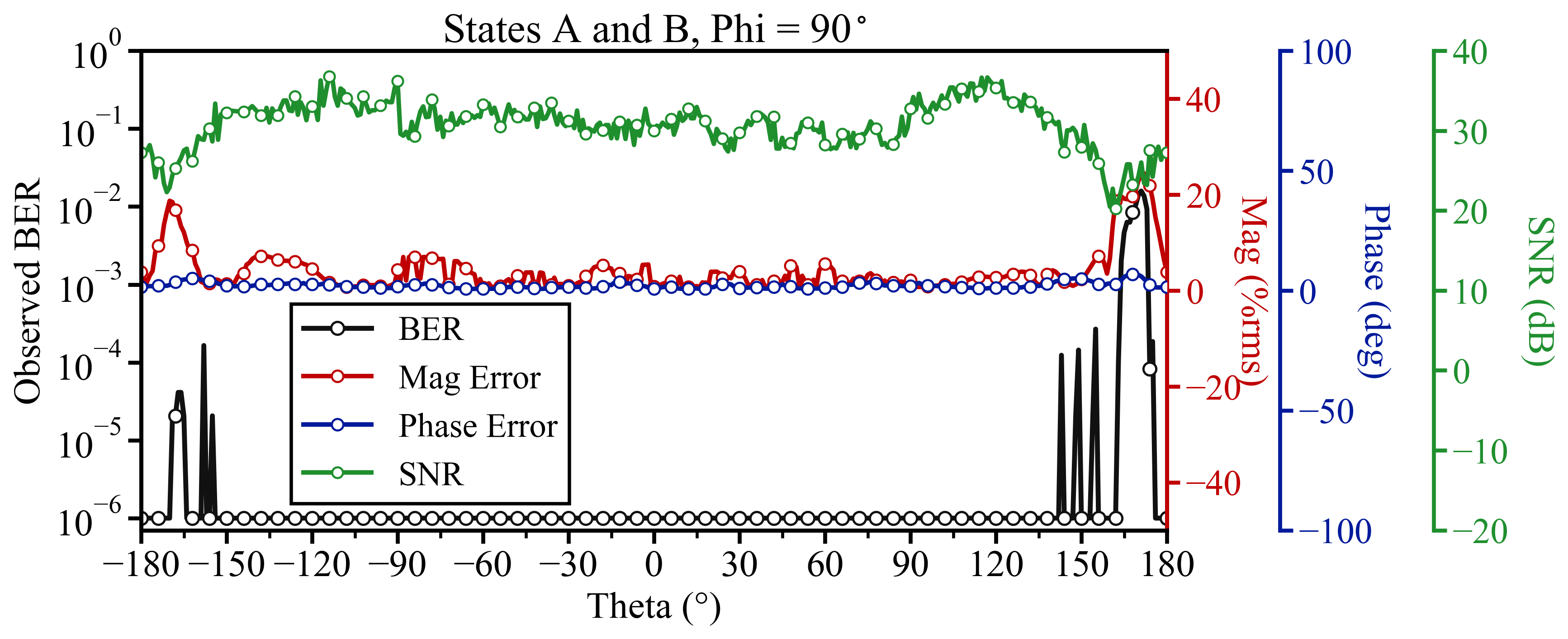}}\hfil\\
\subfloat[]{\includegraphics[width=0.49\textwidth]{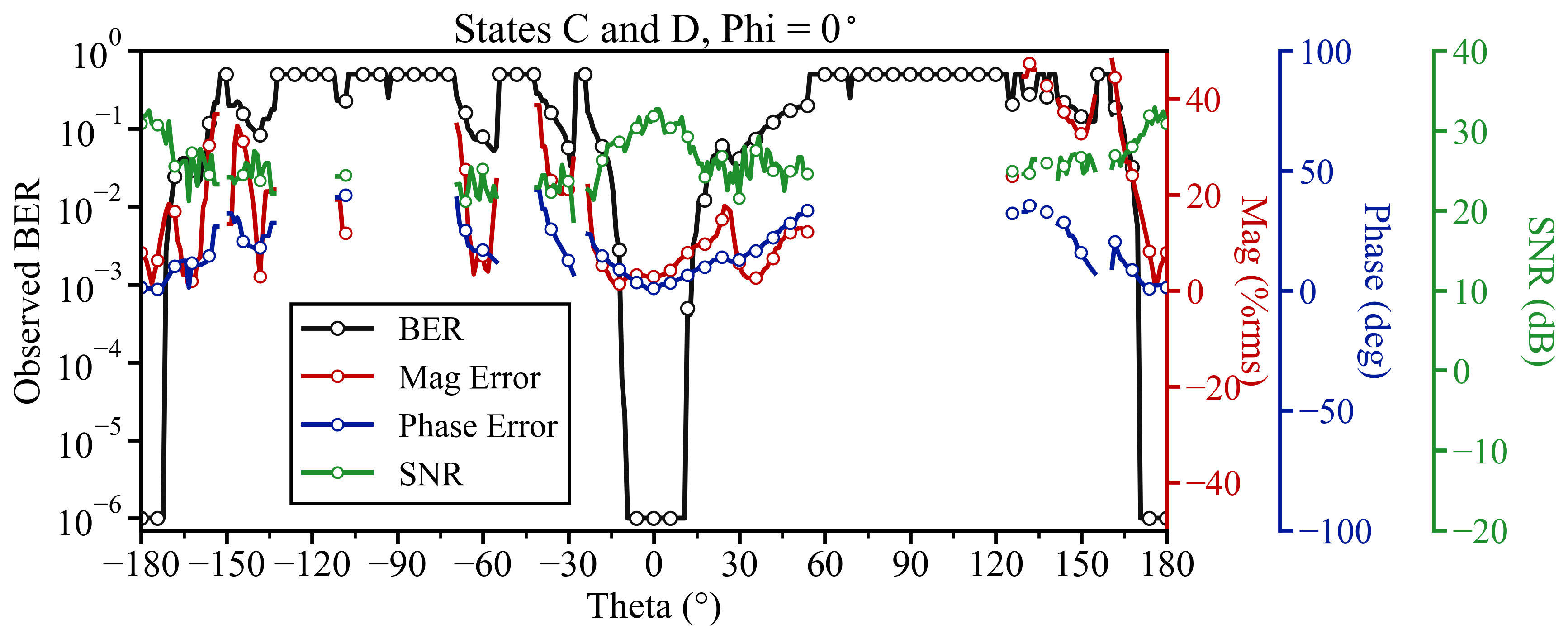}}\hfil
\subfloat[]{\includegraphics[width=0.49\textwidth]{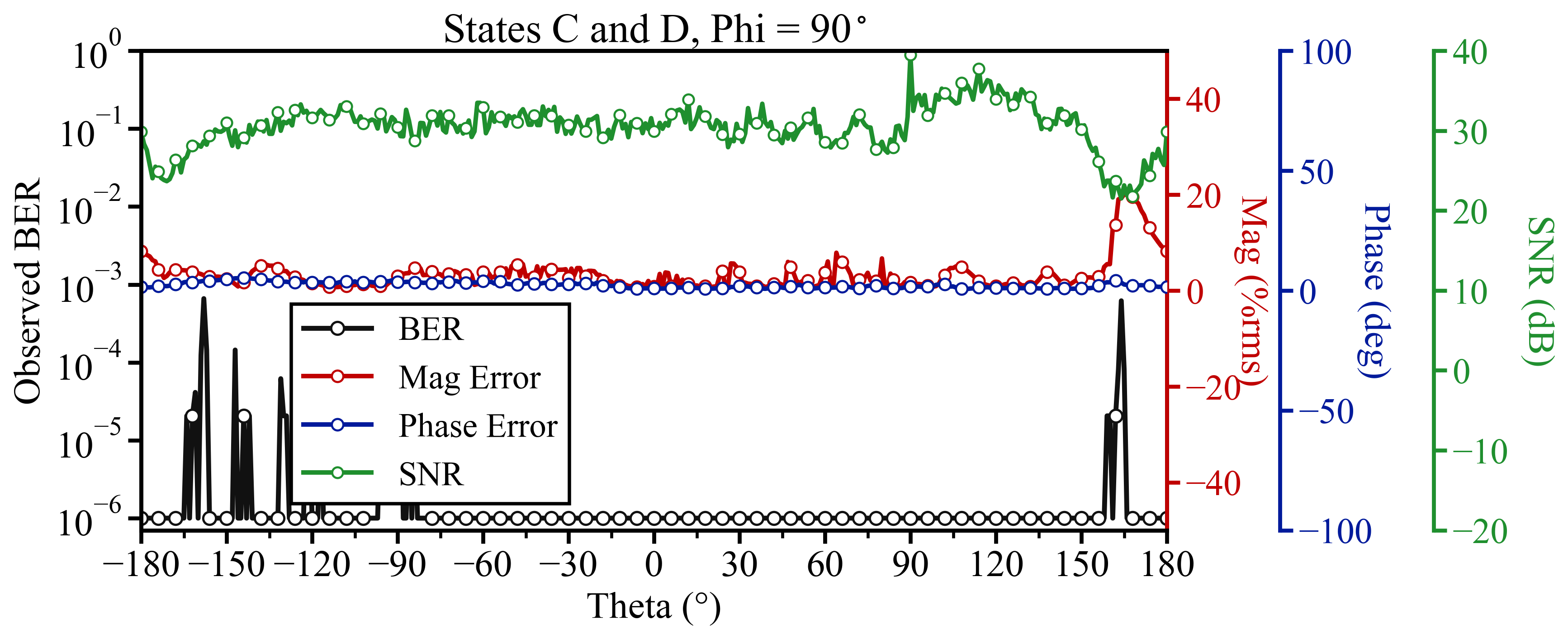}}\hfil\\
\subfloat[]{\includegraphics[width=0.49\textwidth]{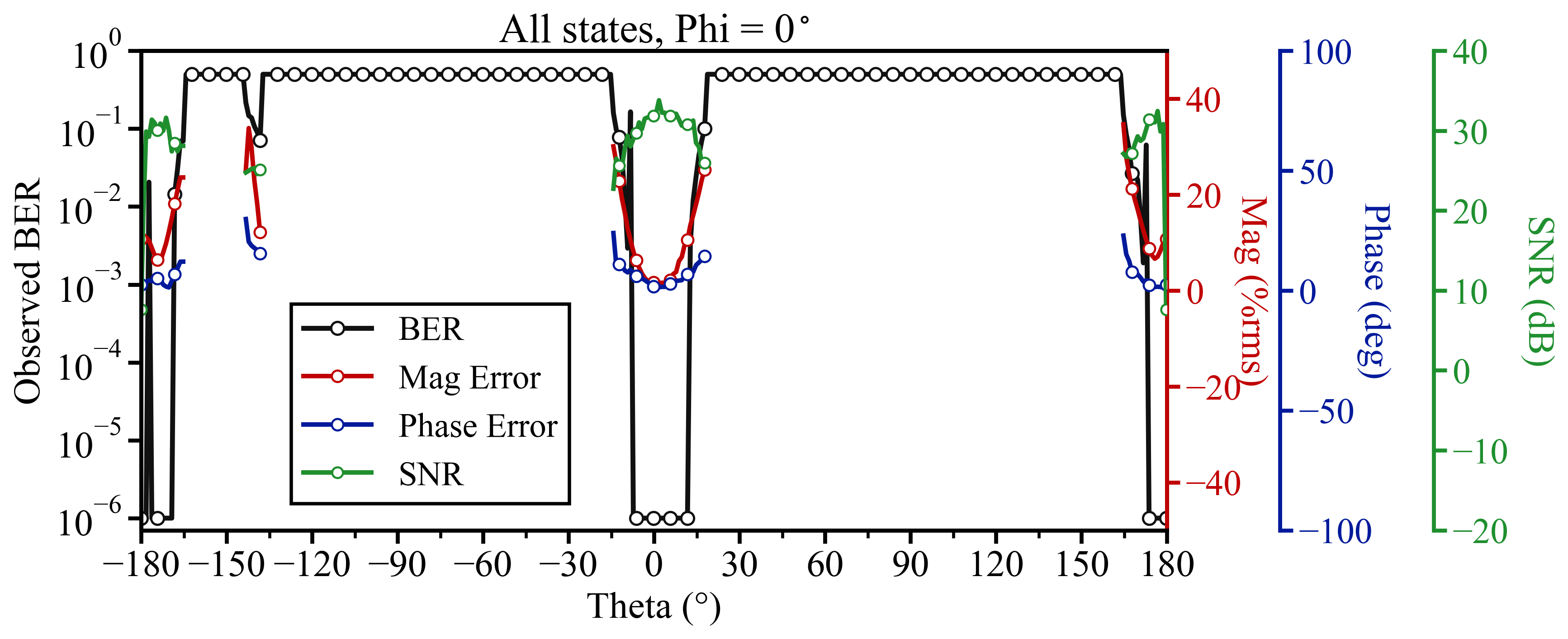}}\hfil
\subfloat[]{\includegraphics[width=0.49\textwidth]{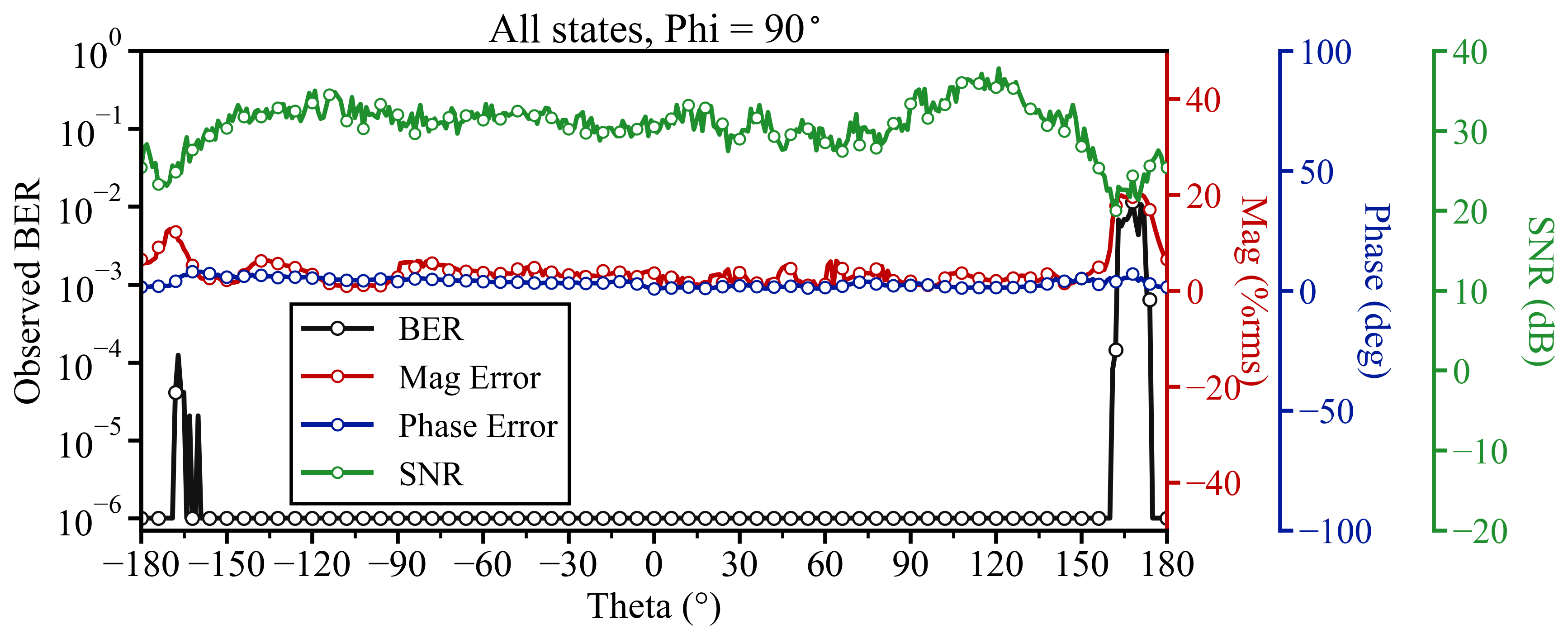}}\hfil
	\caption{Measured communication performance of the proposed dynamic array at 5.05~GHz using 16-QAM. The observed BER, magnitude error, phase error, and received SNR versus $\theta$ are presented for (a) and (b) switching between states A and B in the E- and H-planes, respectively; (c) and (d) switching between states C and D; and (e) and (f) full four-state switching in the E- and H-planes, respectively. Invalid no-demodulation samples are plotted at BER = 0.5 to indicate outage conditions; the corresponding magnitude and phase errors are not shown because they could not be reliably measured when the signal analyzer failed to demodulate the waveform. BER values reported as zero by the signal analyzer are plotted at $10^{-6}$ so that they remain visible on the logarithmic BER axis. Using the BER $\leq 10^{-3}$ criterion, the measured E-plane information beamwidths are $24^\circ$, $21^\circ$, and $19^\circ$ for A--B, C--D, and full four-state switching, respectively.}
	\label{fig:measured_dynamic_ber}
\end{figure*}
\section{Experimental Communication Performance}
The experimental configuration used to characterize the BER performance of the proposed array is illustrated in Fig.~10. The four-element array was fabricated using a standard chemical etching process on a printed circuit substrate. Each antenna element was fed through a coaxial interface using a DC to 8~GHz MHF4-RP jumper cable which was connected to the microstrip feed via a surface-mounted receptacle compatible with the I-PEX MHF4 standard. All interconnections were designed to maintain a characteristic impedance of 50~$\Omega$. 
The dynamic array and its associated switching circuitry were placed at the transmitter. The measured reflection coefficients of the fabricated array are presented in Fig.~\ref{fig:measured_s11}, where the non-excited ports were terminated with matched 50~$\Omega$ loads. A digitally modulated signal with a carrier frequency of 5.05~GHz was generated using a Keysight M8190A arbitrary waveform generator and amplified to ensure operation in a high-SNR regime. At the receiver, two broadband Vivaldi antennas (TSA800), covering 0.8--6~GHz, were used to capture the transmitted signal from the front and back hemispheres, and their RF paths were controlled by an SPDT switch (RF Lambda RFSP2TR0218G) to support full-angle measurement. To enable angular BER characterization, the transmitting array was mounted on a precision rotatable platform, allowing continuous azimuthal scanning while maintaining a fixed far-field separation.

The four-path switching architecture used to implement dynamic excitation is shown in Fig.~10(c). Two double-pole double-throw (DPDT) RF switches (Skyworks SKY13411-374LF) were used to independently control the excitation states of switching pairs. These switches were digitally driven by a Teensy~4.1 microcontroller unit (MCU) to enable real-time path reconfiguration. Digitally controlled phase shifters were inserted in three of the four paths to compensate for phase offsets introduced by cables, attenuators, and other RF components, and also to enable controlled adjustment of the information beam direction. Based on the available laboratory equipment, the output ports of the switching system were calibrated for a differential power within 11 dB to 12.5 dB at 5.05 GHz by using two 6 dB attenuators in two RF paths.

\begin{figure*}[t]
	\centering
\subfloat[]{\includegraphics[width=0.49\textwidth]{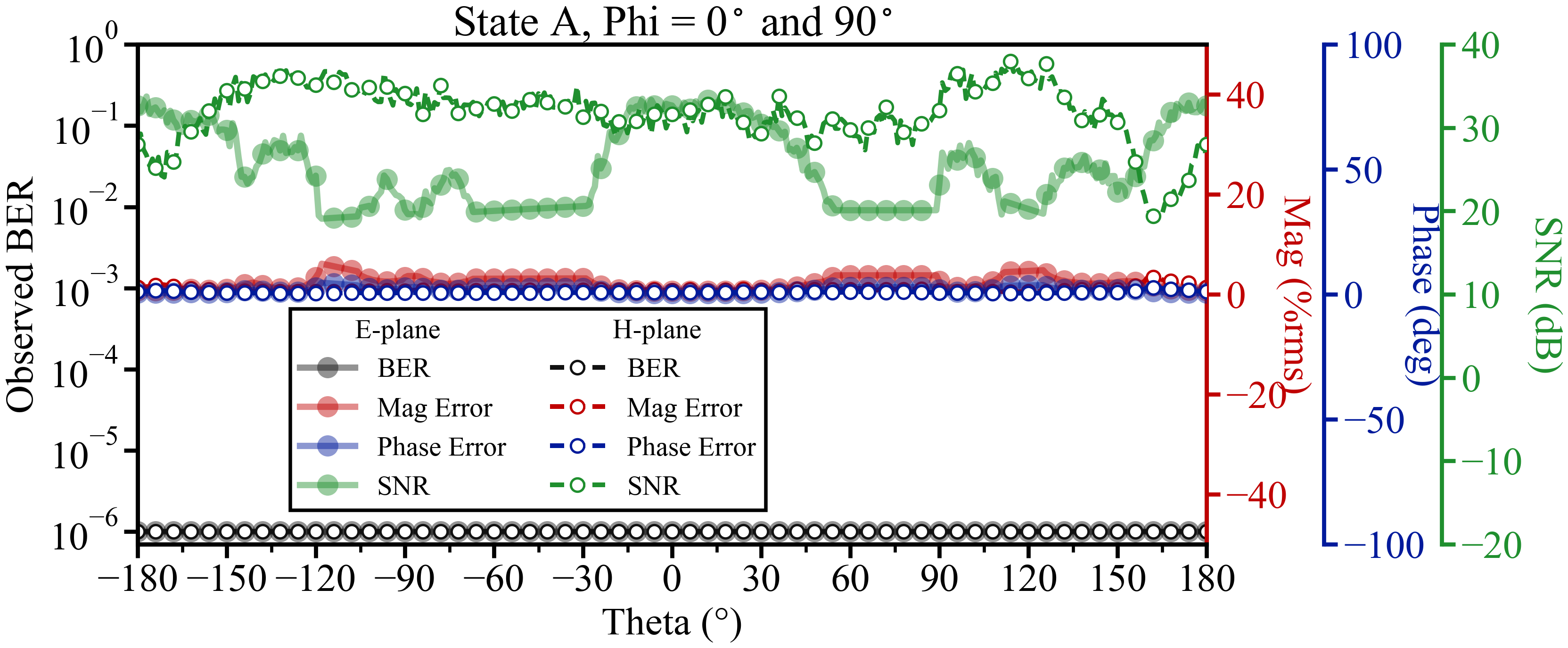}}\hfil
\subfloat[]{\includegraphics[width=0.49\textwidth]{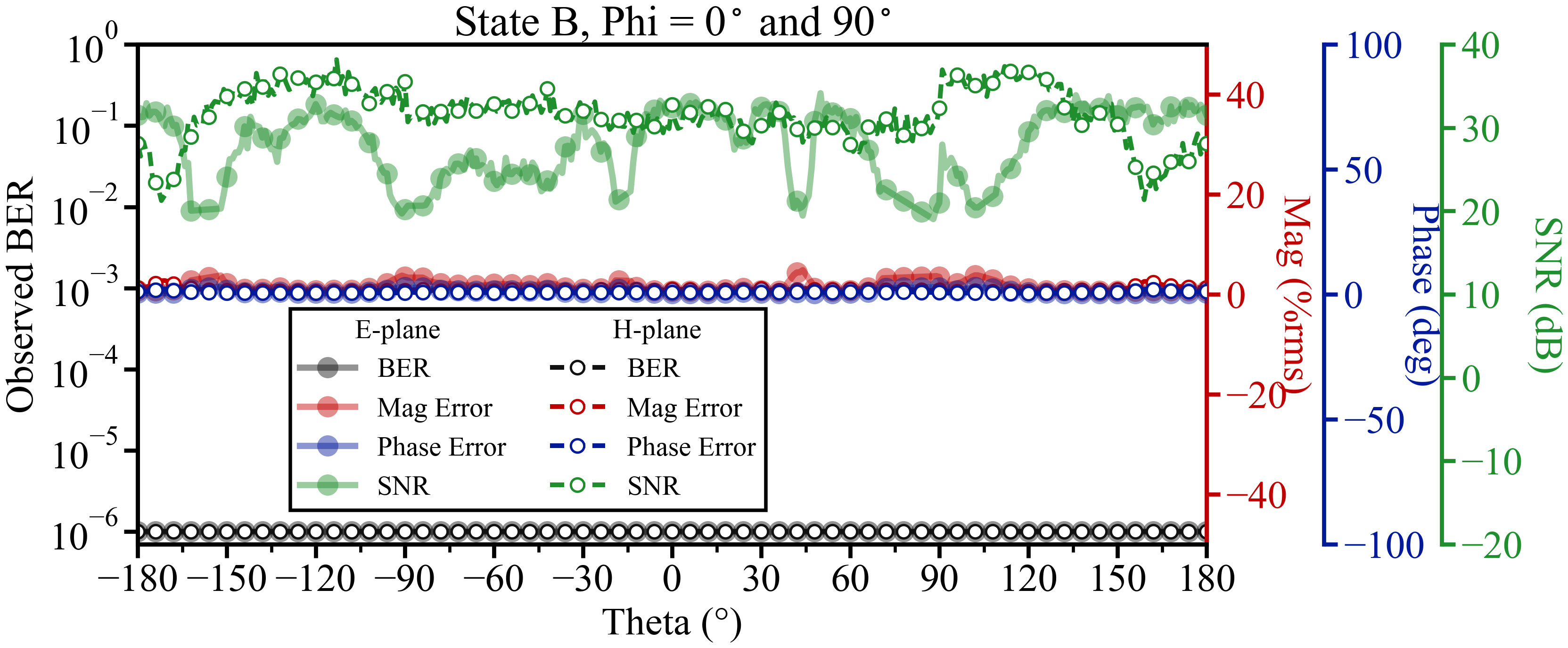}}\\
\subfloat[]{\includegraphics[width=0.49\textwidth]{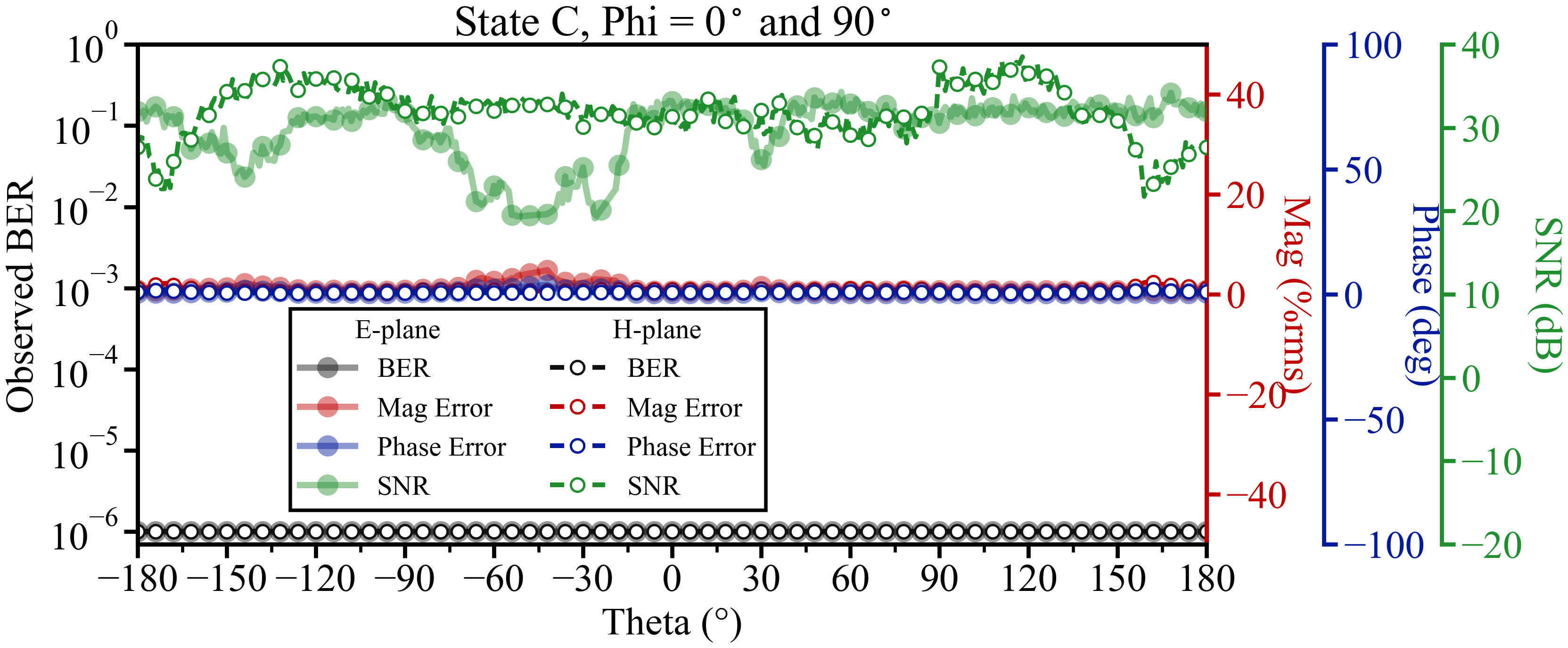}}\hfil
\subfloat[]{\includegraphics[width=0.49\textwidth]{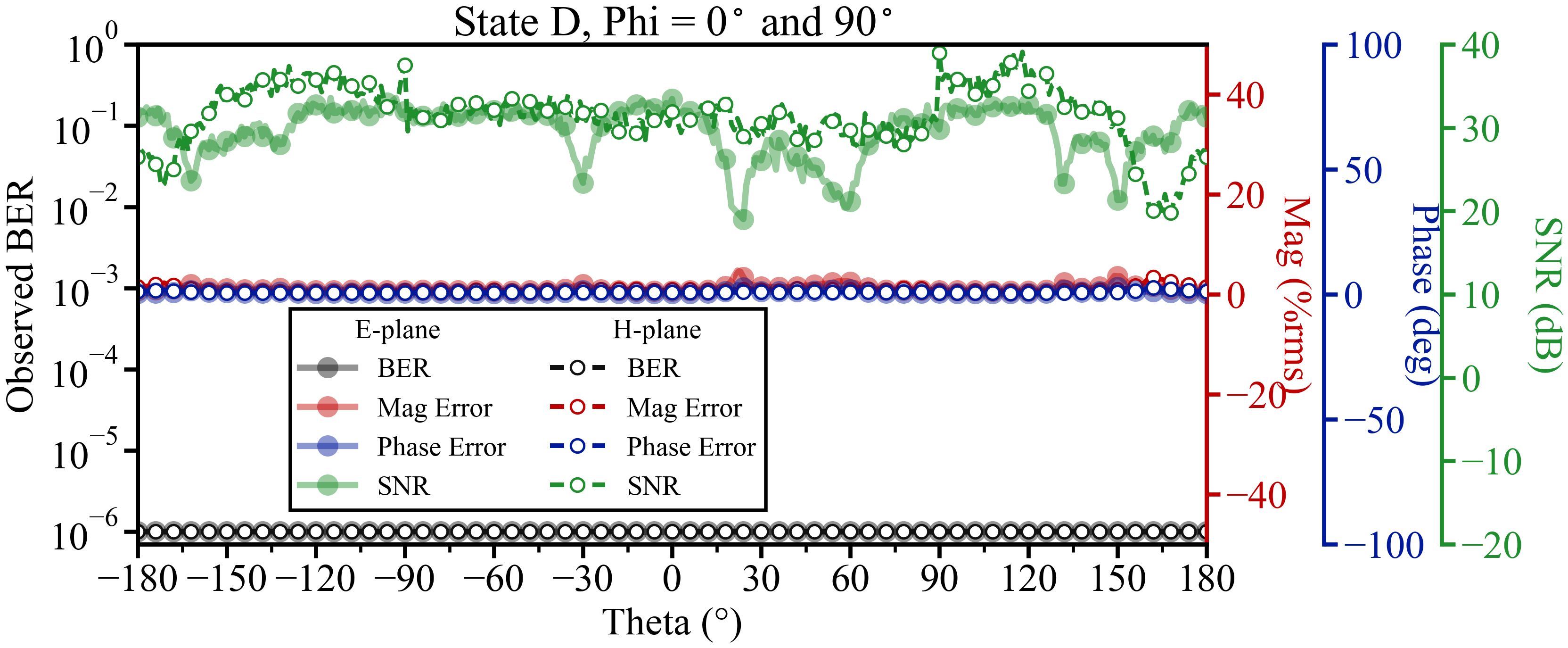}}
	\caption{Measured communication performance of the proposed array under static single-state excitation at 5.05~GHz using 16-QAM. The observed BER, magnitude error, phase error, and received SNR versus $\theta$ are shown with the E-plane and H-plane results overlaid in each static state: (a) State A, (b) State B, (c) State C, and (d) State D. Solid and dashed traces denote the E-plane and H-plane results, respectively. Compared with the dynamic switching results in Fig.~\ref{fig:measured_dynamic_ber}, the static-state cases provide a reference response and demonstrate that the enhanced information-security behavior is introduced by the switching-induced magnitude and phase perturbations.}
	\label{fig:measured_static_ber}
\end{figure*}
The array was operated in transmit mode, and the modulated signal was routed through a broadband four-way power splitter to the switching network. QAM modulation and demodulation were performed using Keysight IQtools and PathWave vector signal analysis software. High-SNR operation was ensured through RF amplification, such that the measured BER variations were dominated by the intentionally manipulated radiation characteristics. Prior to each measurement, system calibration was performed by adjusting the phase shifters and receiver position to minimize magnitude and phase errors at broadside, ensuring a consistent reference condition for angular BER evaluation.

The measured BER, magnitude error, phase error, and received SNR versus elevation angle $\theta$ for all three switching configurations are presented in Fig.~\ref{fig:measured_dynamic_ber}. In the E-plane, all three switching cases exhibit strong angular selectivity consistent with the switching-induced differential radiation predicted in Section~II. In several E-plane angular regions, the switching-induced magnitude and phase errors were sufficiently large that the signal analyzer could not reliably synchronize to and demodulate the received waveform. These invalid no-demodulation samples are plotted at BER = 0.5 in Fig.~\ref{fig:measured_dynamic_ber} to indicate outage conditions, rather than omitted low-BER data, and they represent a more severe form of waveform distortion than the plotted high-BER samples. For A--B switching (Fig.~\ref{fig:measured_dynamic_ber}(a)), the BER decreases sharply to below $10^{-3}$ at broadside with near-zero magnitude and phase errors, confirming reliable information recovery in the intended direction. At off-broadside angles, the magnitude error increases substantially and the BER rises above $10^{-1}$ over most of the angular range, consistent with the dominant real differential component of $AF_{\Delta,AB}$. The C--D switching case (Fig.~\ref{fig:measured_dynamic_ber}(c)) produces a narrower low-BER region at broadside driven primarily by phase perturbation, while the full four-state switching (Fig.~\ref{fig:measured_dynamic_ber}(e)) yields the most confined information beam, achieving an information beamwidth below $25^\circ$ and maintaining elevated BER exceeding $10^{-3}$ throughout the remaining E-plane directions. In all cases, the received SNR remains above 19~dB across the full angular scan, confirming that the observed BER degradation arises from intentional switching-induced waveform distortion rather than insufficient link budget. The H-plane results (Figs.~\ref{fig:measured_dynamic_ber}(b), (d), and (f)) show that the BER remains below $10^{-3}$ over the full $\pm180^\circ$ azimuthal range with consistent SNR and negligible magnitude and phase errors across all switching configurations, experimentally validating the omnidirectional communication coverage in the orthogonal plane.
	
	To isolate the contribution of the dynamic switching mechanism, 
	single-state measurements were performed under each of the four static 
	excitation states, with the E-plane and H-plane results overlaid in 
	Fig.~\ref{fig:measured_static_ber}. Under static excitation, neither 
	plane exhibits any meaningful angular variation in BER, which remains 
	uniformly near $10^{-3}$ throughout the full scan range, with negligible 
	magnitude and phase errors in both planes. This confirms that a fixed 
	excitation state does not introduce sufficient differential radiation to 
	produce spatial information selectivity. The contrast between 
	Figs.~\ref{fig:measured_dynamic_ber} and~\ref{fig:measured_static_ber} 
	directly demonstrates that the angle-dependent BER characteristics 
	observed under dynamic operation originate exclusively from the 
	switching-induced differential component $AF_\Delta$, as described by 
	the average--differential array factor formulation in Section~II. These 
	static-state results thus serve as a controlled baseline confirming that 
	the dynamic switching architecture is essential for realizing effective 
	antenna-level directional modulation.

\section{Conclusion}
This paper demonstrated a compact dynamic antenna array for planar physical-layer security based on antenna-level directional modulation while preserving omnidirectional H-plane coverage. Experimental results using 16-QAM under high-SNR conditions show that reliable communication is confined to a narrow broadside region, while the BER degrades significantly at off-broadside angles, confirming strongly angle-dependent communication performance. The observed behavior is enabled by switching-induced differential magnitude and phase modulation within a four-element planar array of meander-line monopole antennas. While the E-plane exhibits strong waveform distortion outside the intended direction, the H-plane radiation remains quasi-static and omnidirectional, preserving full angular coverage in the orthogonal plane. An average--differential array-factor formulation explains how different switching pairs introduce complementary security mechanisms through their impact on the angular distribution of recoverable information. The antenna is realized on a single-layer commercial substrate with a compact footprint and validated using a low-complexity four-path switching network composed of commercial RF components. These results demonstrate a practical and hardware-efficient dynamic array architecture for secure communication with omnidirectional H-plane coverage and suggest its applicability as a building block for future wireless communications networks.

\bibliographystyle{ieeetr}
\bibliography{areferences}

\end{document}